%% file: conference_101719.tex
\documentclass[conference]{IEEEtran}
\IEEEoverridecommandlockouts
\usepackage{cite}
\usepackage{amsmath,amssymb,amsfonts}
\usepackage{algorithm}
\usepackage{algorithmicx}
\usepackage{algpseudocode}
\usepackage{graphicx}
\usepackage{booktabs}
\usepackage{textcomp}
\usepackage{xcolor}
\usepackage{hyperref}
\usepackage[nolist]{acronym}
\usepackage{paralist}
\def\BibTeX{{\rm B\kern-.05em{\sc i\kern-.025em b}\kern-.08em
    T\kern-.1667em\lower.7ex\hbox{E}\kern-.125emX}}

\begin{document}

\input{acronyms}

\bstctlcite{IEEEexample:BSTcontrol}

\title{Simulation of Multi-Stage Attack and Defense Mechanisms in Smart Grids}

\author{
\IEEEauthorblockN{%
Ömer Sen\IEEEauthorrefmark{1}\IEEEauthorrefmark{2},
Bozhidar Ivanov\IEEEauthorrefmark{1},
Christian Kloos\IEEEauthorrefmark{1},
Christoph Zöll\IEEEauthorrefmark{1},
Philipp Lutat\IEEEauthorrefmark{1},
Martin Henze \IEEEauthorrefmark{3}\IEEEauthorrefmark{4},
Andreas Ulbig \IEEEauthorrefmark{1}\IEEEauthorrefmark{2}, \\
Michael Andres \IEEEauthorrefmark{2},
}

\IEEEauthorblockA{%
\IEEEauthorrefmark{1}\textit{IAEW, RWTH Aachen University,} Aachen, Germany \&
\IEEEauthorrefmark{2}\textit{DE, Fraunhofer FIT} Aachen, Germany\\
Email: \{oemer.sen, andreas.ulbig, michael.andres\}@fit.fraunhofer.de, \{o.sen, p.lutat, a.ulbig\}@iaew.rwth-aachen.de,\\
{bozhidar.ivanov, christoph.zoell, christian.kloos}@rwth-aachen.de}
\IEEEauthorblockA{%
\IEEEauthorrefmark{3}\textit{SPICe, RWTH Aachen University,} Aachen, Germany \&
\IEEEauthorrefmark{4}\textit{CA\&D, Fraunhofer FKIE,} Wachtberg, Germany\\
Email: henze@spice.rwth-aachen.de}
}

\maketitle

\begin{abstract}
The power grid is a vital infrastructure in modern society, essential for ensuring public safety and welfare. As it increasingly relies on digital technologies for its operation, it becomes more vulnerable to sophisticated cyber threats. These threats, if successful, could disrupt the grid’s functionality, leading to severe consequences. To mitigate these risks, it is crucial to develop effective protective measures, such as intrusion detection systems and decision support systems, that can detect and respond to cyber attacks. Machine learning methods have shown great promise in this area, but their effectiveness is often limited by the scarcity of high-quality data, primarily due to confidentiality and access issues.

In response to this challenge, our work introduces an advanced simulation environment that replicates the power grid’s infrastructure and communication behavior. This environment enables the simulation of complex, multi-stage cyber attacks and defensive mechanisms, using attack trees to map the attacker’s steps and a game-theoretic approach to model the defender’s response strategies. The primary goal of this simulation framework is to generate a diverse range of realistic attack data that can be used to train machine learning algorithms for detecting and mitigating cyber attacks. Additionally, the environment supports the evaluation of new security technologies, including advanced decision support systems, by providing a controlled and flexible testing platform.

Our simulation environment is designed to be modular and scalable, supporting the integration of new use cases and attack scenarios without relying heavily on external components. It enables the entire process of scenario generation, data modeling, data point mapping, and power flow simulation, along with the depiction of communication traffic, in a coherent process chain. This ensures that all relevant data needed for cyber security investigations, including the interactions between attacker and defender, are captured under consistent conditions and constraints.

The simulation environment also includes a detailed modeling of communication protocols and grid operation management, providing insights into how attacks propagate through the network. The generated data are validated through laboratory tests, ensuring that the simulation reflects real-world conditions. These datasets are used to train machine learning models for intrusion detection and evaluate their performance, specifically focusing on how well they can detect complex attack patterns in power grid operations.
\end{abstract}


\input{chapter1a}

\input{chapter1b}
\input{chapter1c}
\input{chapter1d}
\input{chapter2a}

\input{chapter2b}
\input{chapter2c}
\input{chapter3}
\input{chapter4d}
\input{chapter4b}
\input{chapter4a}
\input{chapter4c}
\input{chapter5}

\section{Acknowledgments}
\begin{minipage}{0.65\columnwidth}%
This work has received funding from the Federal Ministry of Education and Research (BMBF) under project funding reference 03SF0694A (Beautiful).
\end{minipage}
\hspace{0.02\columnwidth}
\begin{minipage}{0.27\columnwidth}%
	\includegraphics[width=\textwidth]{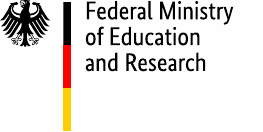}
\end{minipage}

\bibliographystyle{IEEEtran}
\bibliography{conference_101719}

\end{document}

%% file: acronyms.tex
\acrodef{DSO}{Distribution System operator}
\acrodef{PERA}{Purdue Enterprise Reference Architecture}
\acrodef{DMZ}{Demilitarized Zone}
\acrodef{ML}{Machine Learning}
\acrodef{ADT}{Attack-Defense Trees}
\acrodef{DSS}{Decision Support System}
\acrodef{RTT}{Round Trip Time}
\acrodef{C2}{Command and Control}
\acrodef{CPS}{Cyber-Physical System}
\acrodef{VPP}{Virtual Power Plant}
\acrodef{SE}{State Estimation}
\acrodef{BDD}{Bad Data Detection}
\acrodef{OPF}{Optimal Power Flow}
\acrodef{FDI}{False Data Injection}
\acrodef{MITM}{Man-in-the-Middle}
\acrodef{SGAM}{Smart Grid Architecture Model}
\acrodef{ET}{Electrical Energy Technology}
\acrodef{RTU}{Remote Terminal Unit}
\acrodef{IED}{Intelligent Electronic Device}
\acrodef{DoS}{Denial of Service}
\acrodef{MTU}{Master Terminal Unit}
\acrodef{L2}{ISO/OSI Layer 2}
\acrodef{L3}{ISO/OSI Layer 3}
\acrodef{L4}{ISO/OSI Layer 4}
\acrodef{L7}{ISO/OSI Layer 7}
\acrodef{WAN}{Wide Area Network}
\acrodef{SCADA}{Supervisory Control and Data Acquisition}
\acrodef{DER}{Distributed Energy Resources}
\acrodef{ICT}{Information and Communication Technologies}
\acrodef{CIA}{Confidentiality, Integrity and Availability}
\acrodef{ICS}{Industrial Control System}
\acrodef{IT}{Information Technology}
\acrodef{OT}{Operational Technology}
\acrodef{MulVAL}{Multi-host, Multi-stage Vulnerability Analysis}
\acrodef{NVD}{National Vulnerability Database}
\acrodef{OVAL}{Open Vulnerability and Assessment Language}
\acrodef{TTC}{Time-to-Compromise}
\acrodef{IDS}{Intrusion Detection System}
\acrodef{CVE}{Common Vulnerabilities and Exposures}
\acrodef{CVSS}{Common Vulnerability Scoring System}
\acrodef{Ac}{Access Complexity}
\acrodef{Au}{Authentication}
\acrodef{Ex}{Exploitability}
\acrodef{DT}{Decision Tree}
\acrodef{GAN}{Generative Adversarial Network}
\acrodef{RF}{Random Forest}
\acrodef{SVM}{Support Vector Machine}
\acrodef{MCC}{Matthews correlation coefficient}
\acrodef{AUC}{Area Under Curve}
\acrodef{ROC}{Receiver Operating Characteristic}
\acrodef{TP}{True Positive}
\acrodef{TN}{True Negative}
\acrodef{FP}{False Positive}
\acrodef{FN}{False Negative}
\acrodef{CNB}{Complement Naïve Bayes}
\acrodef{XGB}{Extreme Gradient Boosting}
\acrodef{MQTT}{Message Queuing Telemetry Transport}
\acrodef{TTC}{Time-to-Compromise}
\acrodef{ML}{Machine Learning}
\acrodef{HMI}{Human-Machine-Interface}
\acrodef{CI}{Confidence Interval}

%% file: chapter1a.tex
\section{Introduction} \label{sec:introduction}
The rapid modernization of distribution grid infrastructures, driven by renewable energy integration and increased \ac{ICT} reliance, introduces new operational complexities and heightens cyber security vulnerabilities, emphasizing the need for robust management and security frameworks.~\cite{vandervelde2020medit,krause2021cybersecurity}.

The cyber attack on the Ukrainian regional distribution grid serves as an example of the potential disruption caused by such attacks~\cite{b1}.
With the increasing size and complexity of power grids, as well as the growing dependence on digitalization and renewable energy sources, securing cyber security is essential to ensure the reliable functioning of power systems~\cite{lenanrt2023}.

\acp{IDS} are actively developed to enhance the protection of critical infrastructure by observing and scrutinizing network or system behavior to identify potential cyber threats~\cite{b3,wolsing2022ipal}.
Also, \ac{DSS} enhance incident response capabilities by employing advanced techniques such as \ac{ADT} and optimization algorithms, crucial for identifying and countering emerging cyber threats effectively~\cite{zografos2002real, bouramdane2023cyberattacks}.
Detecting and preventing planned attacks on the power system requires the implementation of effective countermeasures, and \ac{ML} algorithms are being developed for this purpose \cite{kus2022false}.
However, the lack of data required to train these algorithms can limit the predictability power of these algorithms, necessitating the generation of synthetic attack data that captures the characteristics of real attacks.

Research on artificial data generation for cyber incidents in critical infrastructure, such as power grids, uses computer simulations to replicate energy and communication systems, enabling realistic attack scenario modeling. These simulations often require specialized models to account for the unique characteristics of smart grid communication protocols and use co-simulations to capture detailed behaviors of \ac{ICT}-based appliances and their associated attack vectors.

For instance, one approach focused on using serious games to simulate attack-defense scenarios, providing valuable insights into strategic decision-making. However, this method lacked flexibility in generating real-time scenarios and adapting to evolving threats \cite{att_dff_1}. Another game-theoretic approach employed for wireless sensor networks offered an optimized defense mechanism against malware propagation but did not adequately represent dynamic attacker behavior in larger, more complex smart grid environments \cite{att_dff_6}. Additionally, a tri-level game-theoretic model for smart grids analyzed interactions at the power plant, transmission, and distribution levels but simplified these interactions, limiting its ability to handle real-time adaptability and scalability \cite{att_dff_7}. Similarly, another work introduced a multi-stage simulation using game theory to optimize the defense of power systems, although it faced challenges in scaling the framework to larger systems \cite{att_dff_11}.

Simulation frameworks and co-simulation environments have also contributed significantly to the field. One such framework integrated red and blue team agents to automate cyber attack and defense simulations but struggled with scalability and real-time adaptability in rapidly evolving threat landscapes \cite{att_dff_2}. A co-simulation framework combining power grid and communication simulators allowed for more detailed analyses but encountered limitations when scaling to larger, real-world grid scenarios and incorporating diverse attack types \cite{att_dff_4}. Another work analyzed structural vulnerabilities in spatial \ac{CPS}, providing a foundational approach for cascading failure simulations. However, it lacked real-time capabilities and failed to simulate the more complex cyber physical interactions critical to smart grid security \cite{att_dff_5}. Furthermore, the tri-level game-theoretic simulation introduced in another study provided insights into strategic resource allocation but was constrained by assumptions of complete information and struggled to model dynamic, real-time behaviors in larger systems \cite{att_dff_11}.

\begin{figure*}
\centerline{\includegraphics[width=\linewidth]{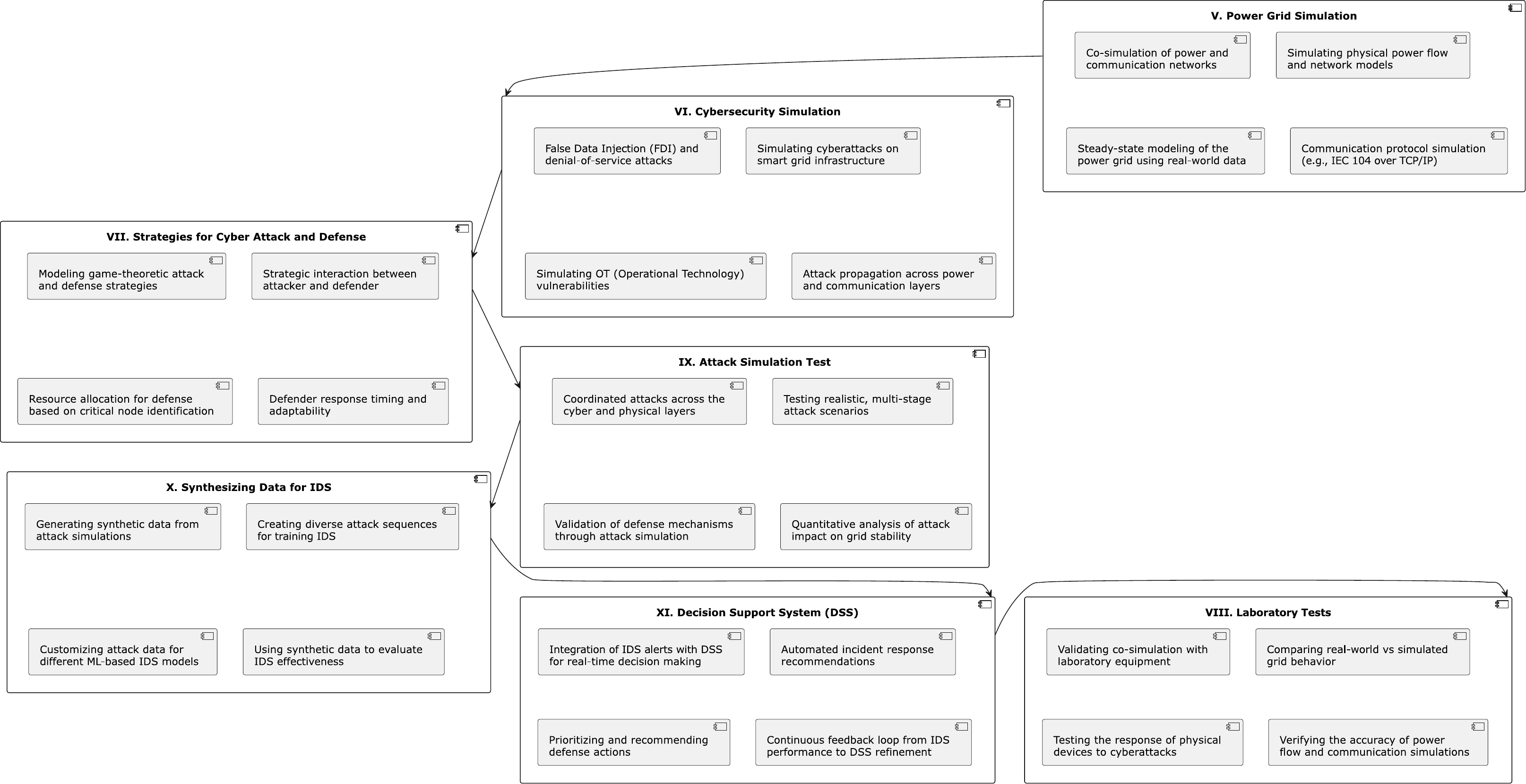}}
\caption{The diagram illustrates the structured methodology of attack and defense simulation for smart grids, highlighting co-simulation, game-theoretic strategies, machine learning-based IDS, and decision support systems, with validation through lab tests.}
\label{fig:framework_overview_structure}
\end{figure*}

In terms of reviewing the current research landscape, one work provided a comprehensive analysis of existing simulation, modeling, and analysis methods for power systems, revealing the limitations of existing frameworks. These include a lack of scalability, limited real-time adaptability, and insufficient integration of complex cyber physical interactions within smart grids \cite{att_dff_9}. This review highlights the need for more advanced simulation environments that can better handle dynamic cyber physical interactions, real-time adaptation to evolving threats, and scalability to larger, interconnected grid systems.

Across these contributions, we identify several research gaps. First, scalability remains a significant challenge, as many of the proposed frameworks struggle to simulate larger, interconnected smart grid systems that mirror real-world complexity. Another key issue is real-time adaptability, with most models lacking the ability to dynamically adjust strategies for both attack and defense as threats evolve. Furthermore, many of the models do not fully capture the dynamic interactions between the cyber and physical components of smart grids, a crucial aspect of these systems. Lastly, most existing frameworks offer limited flexibility in simulating complex, multi-stage attack scenarios and diverse cyber attack vectors, which hinders their ability to provide realistic training data for \ac{ML}-based \ac{IDS}.

This work aims to address these gaps by developing a sophisticated multi-stage attack-defense simulation framework. This framework leverages co-simulation techniques to integrate both the power grid and its communication systems while incorporating game-theoretic principles to model dynamic, real-time strategies. By generating synthetic attack data specifically for \ac{ML}-based \ac{IDS}, the work seeks to enhance the robustness of smart grid cyber security. Additionally, this framework includes multi-stage, dynamic attack models capable of adapting in real-time, thereby filling the gaps in scalability, real-time adaptability, and diversity of attack scenarios identified in prior research.

Figure~\ref{fig:framework_overview_structure} illustrates the attack and defense simulation methodology for smart grids integrating a co-simulation environment that models both the physical power flows and communication protocols (IEC 60870-5-104 over TCP/IP), enabling realistic simulations of cyber attacks such as \ac{FDI} and \ac{DoS} attacks. A game-theoretic approach dynamically models the interaction between attackers and defenders, optimizing defense strategies based on critical node identification and resource allocation. \ac{ML}-based \ac{IDS} are placed strategically and trained with synthetic data generated from the simulations, while a \ac{DSS} automates real-time incident response based on \ac{IDS} alerts. Intermediate validation through lab tests ensures the accuracy of the simulations, particularly for communication protocols and multi-stage attack sequences, enhancing the system's resilience to evolving cyber threats. This research integrates a holistic simulation framework for smart grids, covering both physical power flows and communication layers using co-simulation techniques and standard protocols. The novel closed-loop environment allows for seamless simulation of grid operations under both normal and attack scenarios, ensuring no reliance on external components. Validation is conducted through lab-based experiments, enhancing the accuracy of the results and supporting the development of machine learning-based IDS and Decision Support Systems for improved smart grid resilience.

With this approach, we aim to advance the field of cyber security in smart grids by developing a sophisticated simulation environment and model. This environment is designed to generate synthetic data for \ac{ML}-based \ac{IDS} and \ac{DSS}, reflecting the complex dynamics between attackers and defenders. Our key contributions are:

\begin{enumerate}
    \item Comprehensive analysis of existing literature and identification of gaps in the current research landscape. Our work delves into benchmarking, data synthesis, and simulation environments, leading to a detailed problem analysis relevant to cyber security in power grids (Sections~\ref{sec:background} and~\ref{sec:relatedwork}).
    \item Development of an innovative model for generating synthetic cyber attack data, tailored for \ac{ML}-based \ac{IDS} applications, enhancing the robustness of power grid cyber security (Section~\ref{sec:cssim}). In particular, introduction of a novel method incorporating attack tree modeling combined with game theory mechanics. This approach is designed to yield diverse and realistic attack data sets, crucial for training advanced \ac{ML} algorithms (Section~\ref{sec:framework}).
    \item Verification and validation of the proposed approaches through laboratory tests and simulation studies. This step ensures the accuracy and reliability of our model and methods in real-world scenarios (Sections~\ref{sec:labres} and ~\ref{sec:resatt}).
    \item Rigorous evaluation and investigation of different \ac{ML} models using the synthetic data generated. This includes a thorough examination of the impact of attacker-defender dynamics on detection quality and attack complexity (Sections~\ref{sec:result} and ~\ref{sec:dss}).
\end{enumerate}

A preliminary version of this paper appears in the proceedings of the 2023 International Conference on Smart Energy Systems and Technologies (SEST)~\cite{sen2023investigation}. We have expanded and enhanced our previous work in several significant ways:
\begin{enumerate}
    \item We have introduced a dedicated section for reviewing relevant literature and a discussion of related work in Section~\ref{sec:relatedwork}. This section now offers a more comprehensive and extensive overview of the research landscape in benchmarking, data synthesis, and simulation environments, culminating in a detailed problem analysis.
    \item A formal description of the underlying simulation environment, which replicates the smart grid along with its communication behavior, has been added in Section~\ref{sec:gridsimulation}, while providing a comprehensive overview of the developed framework in Section~\ref{sec:overview}. 
    \item More detailed insights into our methodology for simulating multi-stage cyber attacks are provided, particularly focusing on the phases of \ac{IT} propagation and \ac{FDI} attack coordination. This includes specific details regarding attack modeling and implementation, as discussed in Section~\ref{sec:cssim}.
    \item The presentation of our evaluation results has been extended to demonstrate the efficacy of the generated data in training \ac{ML}-based \ac{IDS}. This includes various experiments to validate the simulation environment through laboratory tests (Section~\ref{sec:labres}), to examine attack propagation behavior (Section~\ref{sec:resatt}), and to showcase additional use cases such as the demonstration of \ac{DSS} (Section~\ref{sec:dss}).
    \item Furthermore, this significantly extended paper now includes more comprehensive details across all aspects of the design, implementation, and evaluation of the simulation environment.
\end{enumerate}

The structure of this paper is organized as follows: Section~\ref{sec:background} introduction into the relevant foundations, while Section~\ref{sec:relatedwork} discusses the research landscape pertinent to this study. Section~\ref{sec:overview} presents the simulation environment, and Section~\ref{sec:gridsimulation} delves into the details of the grid simulation. Our cyber security simulation is described in Section~\ref{sec:cssim}, whereas Section~\ref{sec:framework} elaborates on the methodology used for generating synthetic attack data. Section~\ref{sec:labres} details the verification of the simulation environment through laboratory tests. The attack simulation is examined in Section~\ref{sec:resatt}, followed by the evaluation of the generated data and the \ac{ML} models for \ac{IDS} in Section~\ref{sec:result} and for \ac{DSS} in Section~\ref{sec:dss}. The paper concludes with a discussion of the results in Section~\ref{sec:conclusion}.

%% file: chapter1b.tex
\section{Background} \label{sec:background}
This section delves into the key components and challenges associated with the operation and control of power grids, emphasizing the significance of cyber security in this evolving landscape.
\subsection{Power Grid Control Structure} \label{subsec:background_powergrid}
The operation and control structure of power grids, similarly to \ac{ICS} \cite{henze2017network}, involves different layers of technology, each playing a vital role in ensuring efficient, secure, and reliable electricity distribution. The electrical energy power grids are often structured into multiple voltage levels, interconnected through three-phase transformers~\cite{schavemaker2017electrical}, i.e. transmission and distribution grids. The grid is a composite of primary and secondary technologies, with the primary technology encompassing components directly involved in the generation, transformation, and transport of electrical energy~\cite{amin2008electric}. The secondary technology complements this by monitoring, controlling, and protecting these components.

Remote control technology, essential for communication between control systems and secondary technology, employs \acp{RTU} to collect and relay information such as measurements and control commands over \ac{WAN}~\cite{gharavi2011smart}. These \acp{RTU} form the critical link between control and process levels, managing monitoring and control data. \ac{SCADA} systems, integral to grid control technology, are deployed as monitoring and control systems within electrical power and pipeline grids. These systems are responsible for centralized data collection and visualization. The hierarchical structure of grid control stations, essential for grid operation, differentiates by their functional scope and responsibility areas~\cite{buchholz2014smart}. These stations are categorized into grid, station, and field control levels, with higher-level control stations overseeing the lower levels remotely.

In addition to these systems, higher-level decision and optimization functions are implemented in grid control stations, handling tasks such as power flow calculation, grid condition detection, and grid security assessment~\cite{masters2013renewable}. For effective operation management, power grid operators undertake responsibilities such as grid monitoring and control, disturbance detection and resolution, and executing switching plans for maintenance~\cite{montoya2021heuristic}. To achieve these tasks, operators use tools such as power flow calculation and \ac{OPF}, along with grid condition detection techniques.

The Purdue Model, a key component of the \ac{PERA}~\cite{williams1996overview}, serves as a vital framework for understanding the various layers and components of \ac{ICS} networks, and consequently for the \ac{SCADA} systems in power grids. This model delineates the \ac{ICS} architecture into two primary zones: the \ac{OT} zone, consisting of physical equipment and operational process control systems, and the \ac{IT} zone, which includes systems for data management and communication. Additionally, a \ac{DMZ} is established between the \ac{IT} and \ac{OT} zones to regulate access. The model divides the \ac{ICS} architecture into six distinct levels:
At the base, Level 0 contains the physical hardware such as transformers and circuit-breakers. Level 1 incorporates controllers such as \acp{IED} and \acp{RTU} that manage the Level 0 devices and bridge the layer with the overlaying layers. Ascending to Level 2, we find the supervisory systems, including \ac{HMI} and \ac{SCADA} software, which oversee \ac{ICS} operations. Level 3 is dedicated to managing production workflows, while Level 4 extends to systems handling logistics and data storage. The top tier, Level 5, encompasses the enterprise network, integrating \ac{ICS} data for strategic business decision-making.

Focusing on the communication aspect, the primary facilities link to the \ac{OT} network via \acp{IED} which handles control and measurement operations, consolidating them via Modbus~\cite{23_Modbus_2012} at the level of \acp{RTU}. This data is transmitted to the \ac{SCADA} system over the \ac{OT} network, employing relevant OT protocols such as IEC 60870-5-104~\cite{22_IEC104_2006}. In this network, the \ac{MTU}, functioning as the \ac{RTU}'s counterpart, serves as a conduit to the \ac{SCADA} system.

\subsection{Cybersecurity in Power Grids} \label{subsec:background_cybersecurity}
Cybersecurity in power grids plays a pivotal role in securing electrical power systems against cyber threats. The contemporary power grid, a complex interplay of generation, transmission, and distribution networks, has grown increasingly dependent on digital technologies for communication and control. This shift towards digitalization, although beneficial for operational efficiency, introduces vulnerabilities to cyber threats, impacting the grid's security and reliability.

The grid's interconnected components, including generators, substations, transformers, and consumer interfaces, feature cyber components such as control software and automated management systems. These components are prone to cyber attacks such as malware, which can interrupt power supply, inflict physical damage, or lead to severe failures. The utilization of \ac{SCADA} systems in grid monitoring and control, traditionally focused more on efficiency than security, are especially vulnerable to cyber attacks through exposed interfaces~\cite{krause2021cybersecurity,james2019improving}. Additionally, the essential grid components such as distributed control systems and \acp{IED} constitute significant potential attack vectors. Incidents, including the 2016 Ukrainian power grid attack~\cite{b1}, underscore these vulnerabilities' real-world consequences, such as widespread blackouts and disruption of vital services. Other forms of attacks include \ac{DoS} attacks, which can overload communication networks, and \ac{FDI} that manipulate data, leading to erroneous operational decisions~\cite{rawat2015detection, ramirez2022classifying}. 

In Europe's energy sector, the IEC 60870-5-104 protocol is prevalent for managing widely dispersed processes~\cite{21_matouvsek2017description}. As a legacy protocol, IEC 60870-5-104 lacks crucial security features such as encryption and authentication, making critical traffic vulnerable to interception by unauthorized entities~\cite{30_wg152016iec}. Attackers can exploit this vulnerability through \ac{MITM} attacks or by establishing unauthorized connections to manipulate traffic~\cite{31_yang2012man}. The IEC 62351 standard introduces new security requirements such as secure end-to-end communication using TLS \cite{wagner2024madtls}, offering key exchange, encryption, and authentication~\cite{32_iec62351_2018}. However, implementing these standards in traditional networks faces challenges due to resource-constrained devices, potentially impacting service availability and real-time performance requirements~\cite{8_tanveer2020secure,34_todeschini2020securing}. Despite these advancements, power grids often have assets with limited performance capabilities and long depreciation periods, necessitating solutions that are compliant with legacy systems~\cite{35_castellanos2017legacy}. 

To combat these threats, various strategies are employed, ranging from \acp{IDS} to robust network architecture designs that segment the grid’s communication networks, enhancing resilience against cyber attacks~\cite{cintuglu2016survey, liu2013framework}. Furthermore, ongoing research in artificial intelligence and \ac{ML} presents new opportunities for predictive cyber security measures, potentially identifying and mitigating threats before they materialize. \ac{IDS} serve as a passive security mechanism, detecting potential attack indicators without actively interfering with network operations~\cite{36_fernandes2019comprehensive,37_zuech2015intrusion}.

\subsection{Intrusion Detection Systems}\label{subsec:background_ids}
An \ac{IDS} is an easily retroffitable security solution, designed to monitor and analyze network or system activities to detect potential cyber threats such as unauthorized access, misuse, or disruption \cite{krause2021cybersecurity,wolsing2022ipal}. \acp{IDS} can be implemented in various forms, including on individual devices or across a network, and they are capable of monitoring a wide range of data types, including device logs, network traffic, and process data, to identify suspicious or malicious activities.

\acp{IDS} are often categorized based on the source of their data into network-based and host-based systems. Host-based \acp{IDS} operate directly on devices, scanning for malicious behavior such as file modifications~\cite{hu2009simple}, while network-based \acp{IDS} analyze network traffic, such as packets or flow-data, to detect malicious activities~\cite{ring2019survey}. This distinction is particularly relevant in \ac{ICS} environments, such as power grids, where network-based \acp{IDS} for \ac{OT} are critical. These systems can recognize and analyze process data, including measurements or control commands, and detect anomalies at the process level.

\ac{ML}-based \acp{IDS} employ sophisticated methods for detecting anomalies \cite{kus2022false}. Two primary approaches used are allowlisting and blocklisting. Allowlisting focuses on identifying deviations from normal system behavior learned by the model, whereas blocklisting identifies known abnormal behaviors represented in the trained model. In both approaches, anomalies may indicate cyber attacks, system malfunctions, or misconfigurations, manifesting as protocol errors, missing messages, communication aborts, and more. These methods leverage network structure, system discovery, and connection behavior to validate the legitimacy of devices, protocols, and communication patterns.

Evaluation metrics for \acp{IDS} are application-specific, with high detection accuracy, runtime efficiency, and explainability being key criteria \cite{lamberts2023evaluations}. The robustness of \acp{IDS} is essential for improved detection accuracy, and the diversity in ensemble methods is a critical aspect to be assessed~\cite{liu2019machine, zhou2012ensemble}. The evaluation typically involves utilizing multiple datasets~\cite{boenninghausen2024commids}, incorporating statistical tests and confidence intervals to ensure reproducibility, and controlling randomization in model evaluations.

%% file: chapter1c.tex
\section{Related Work}\label{sec:relatedwork}
This section reviews relevant literature and studies in the field of cybersecurity for smart grids, highlighting their key advances and identifying areas for further research.

\subsection{Benchmarking} \label{subsec:relatedwork_benchmarking}
The research landscape has been actively exploring the development of benchmark environments to evaluate the performance of various \ac{ML} algorithms in detecting anomalies and intrusions within \ac{ICS}. Approaches such as the Penn \ac{ML} Benchmark~\cite{olson2017pmlb} and the Scientific \ac{ML} Benchmark suite~\cite{thiyagalingam2022scientific} provide vital resources for testing and comparing algorithm performance. Researchers have also emphasized the creation of datasets from real \ac{ICS} environments, such as the Cyber-kit datasets~\cite{mubarak2021ics} and the Numenta anomaly benchmark~\cite{lavin2015evaluating}, to evaluate unsupervised anomaly detection techniques or \acp{IDS} in \ac{ICS}. Specific focus has been given to datasets related to power grids to assess the performance of \ac{ML} algorithms in detecting anomalies in these systems~\cite{bernieri2019evaluation, liyakkathali2020validating, mohammadpourfard2019benchmark}. However, there exists a need for more specialized and comprehensive approaches, particularly in considering the unique characteristics and constraints of \ac{ICS} and power grids~\cite{japkowicz2006question, tufan2021anomaly}.

In addition, research works have investigated \acp{IDS} in \ac{ICS}, focusing on evaluating the criticality of \ac{ICS} devices and identifying potential adversary traces~\cite{15_cook2017industrial, 16_escudero2018process}. Advanced approaches include the replication of program states in digital twins~\cite{9_eckhart2018specification}, cyber attack classification~\cite{18_mohan2020distributed}, and modeling \ac{ICS}/\ac{SCADA} communication using probabilistic automata~\cite{19_matouvsek2021efficient, 20_almseidin2019fuzzy}. Anomaly detection methods for IEC 60870-5-104 have also been explored using multivariate access control and outlier detection approaches~\cite{grammatikis2020anomaly, burgetova2021anomaly, anwar2021comparison}. However, these proposed approaches require additional analytical resources for their functionality, such as infrastructure specifications, attack target understanding, statistical data, or technical specifications~\cite{scheben2017status, dang2021improving, holzinger2020measuring}. 

\subsection{Dataset Generation} \label{subsec:relatedwork_dataset}
A wide range of approaches has been explored for generating synthetic cyber attack data for power grids, including the creation of lab environments such as the CICIDS2017 dataset~\cite{sharafaldin2018toward}, the development of the ID2T framework for reproducible datasets~\cite{cordero2021generating}, modification of existing datasets such as UNSW-NB15 using denoising autoencoders and Wasserstein \acp{GAN}~\cite{pandey2021gan}, and the generation of artificial attack data for power grid security using frameworks such as Melody~\cite{b4}.
Furthermore, studies such as~\cite{dutta2023deep} demonstrate the potential of data-driven deep reinforcement learning for proactive cyber defense, while~\cite{agnew2022implementation} introduces a cross-layered framework for securing the power grid.

Generating comprehensive and realistic datasets for effective \ac{ML} model training is challenging due to various factors such as infrastructure implementation, scenario development, and ensuring data integrity and privacy \cite{uetz2021socbed}.
Publicly available datasets may be limited \cite{mitseva2022challenges,lamberts2023evaluations} as well as sharing real-world data publicly can increase the risk of cyber attacks, highlighting the need for more artificial datasets. 

\subsection{Simulation Environment} \label{subsec:relatedwork_simulation}
Simulation environments are crucial for advancing cyber security in power grids, allowing for the safe testing and refinement of cyber security strategies. These environments have evolved to cater to diverse cyber security scenarios in the power grid sector. 

Co-simulation platforms have become increasingly significant, integrating both the cyber and physical aspects of power grids. These platforms are categorized into hardware-based and software-based simulations. Hardware-based simulations, such as those using real-time platforms such as OPAL-RT~\cite{gomez2019real}, RTDS~\cite{sharma2015testing}, and Typhoon-HiL~\cite{jia2020real}, offer high-fidelity modeling essential for real-time response analysis. These platforms are crucial for scenarios where interaction between physical hardware components and simulated systems is needed. In contrast, software-based co-simulation platforms utilize tools such as mosaik~\cite{rueda2022comparison}, OMNeT++~\cite{troiano2016co}, and ns-3~\cite{tariq2014cyber} to provide flexible and scalable environments. These platforms are adept at modeling complex network interactions and cyber security protocols in a purely virtual setup, making them versatile for various research contexts. 

The development of a power grid application platform exemplifies the effectiveness of these advanced co-simulation environments in analyzing the interplay between power systems and \ac{ICT}~\cite{amarasekara2015co}. Network and power grid co-simulation frameworks provide valuable insights into grid vulnerabilities and cyber security readiness, enhancing our understanding of wide-area monitoring networks~\cite{bhor2016network}. Platforms such as DSSnet, which combine electrical power distribution system simulation with software-defined networking emulation, are examples of software-based simulations demonstrating potential in power grid planning and evaluation~\cite{hannon2018combining}. In contrast, hardware-based platforms facilitate transient state simulations that provide detailed insights into the dynamic behavior of power grids under cyber attack scenarios, crucial for developing effective mitigation strategies against evolving cyber threats~\cite{bian2015real}. Testbeds such as PowerCyber, representing hardware-based simulations, provide realistic environments for testing and validating security solutions and are invaluable for practical, hands-on cyber security research~\cite{hahn2010development}. 

The integration of power and communication network simulations for power grid applications~\cite{hopkinson2006epochs} highlights the significance of combining both power approaches in cybersecurity research.

\subsection{Problem Statement} \label{subsec:relatedwork_problem}
The necessity for an abstract simulation environment dedicated to cyber security research in power grids, particularly for exploring defense mechanisms such as \ac{ML}-based \acp{IDS} and \acp{DSS}, is evident from the existing gaps in co-simulation approaches. Current models, while capable of cross-domain research, often fall short in focusing specifically on cyber security. 

This shortcoming underscores the importance of a simulation environment that can accurately mimic real-world cyber attacks, a critical factor for testing and enhancing defense systems. Such an environment is paramount for generating comprehensive and realistic datasets essential for training \ac{ML} algorithms in detecting cyber threats. The richness and diversity of these datasets directly influence the effectiveness of \ac{ML}-based security solutions. Additionally, \acp{DSS}, reliant on accurate and extensive data, would benefit significantly from a simulation environment that replicates real-world scenarios with precision. 

One of the critical challenges faced by existing approaches is their limitations in terms of scalability and flexibility. An advanced simulation environment needs to overcome these hurdles, allowing researchers to expand the complexity of simulations and alter scenarios as necessary, which is vital for in-depth cyber security research.

Moreover, reducing the dependency on expensive, proprietary hardware is also crucial. By leveraging advanced software-based simulations, such an environment could make cyber security research more accessible and practical. Furthermore, a comprehensive cross-domain analysis is essential. The simulation environment should consider both the communication and power aspects of power grids to understand the interdependencies thoroughly and develop robust cyber security strategies.

%% file: chapter1d.tex
\begin{figure*}\
\centerline{\includegraphics[width=\linewidth]{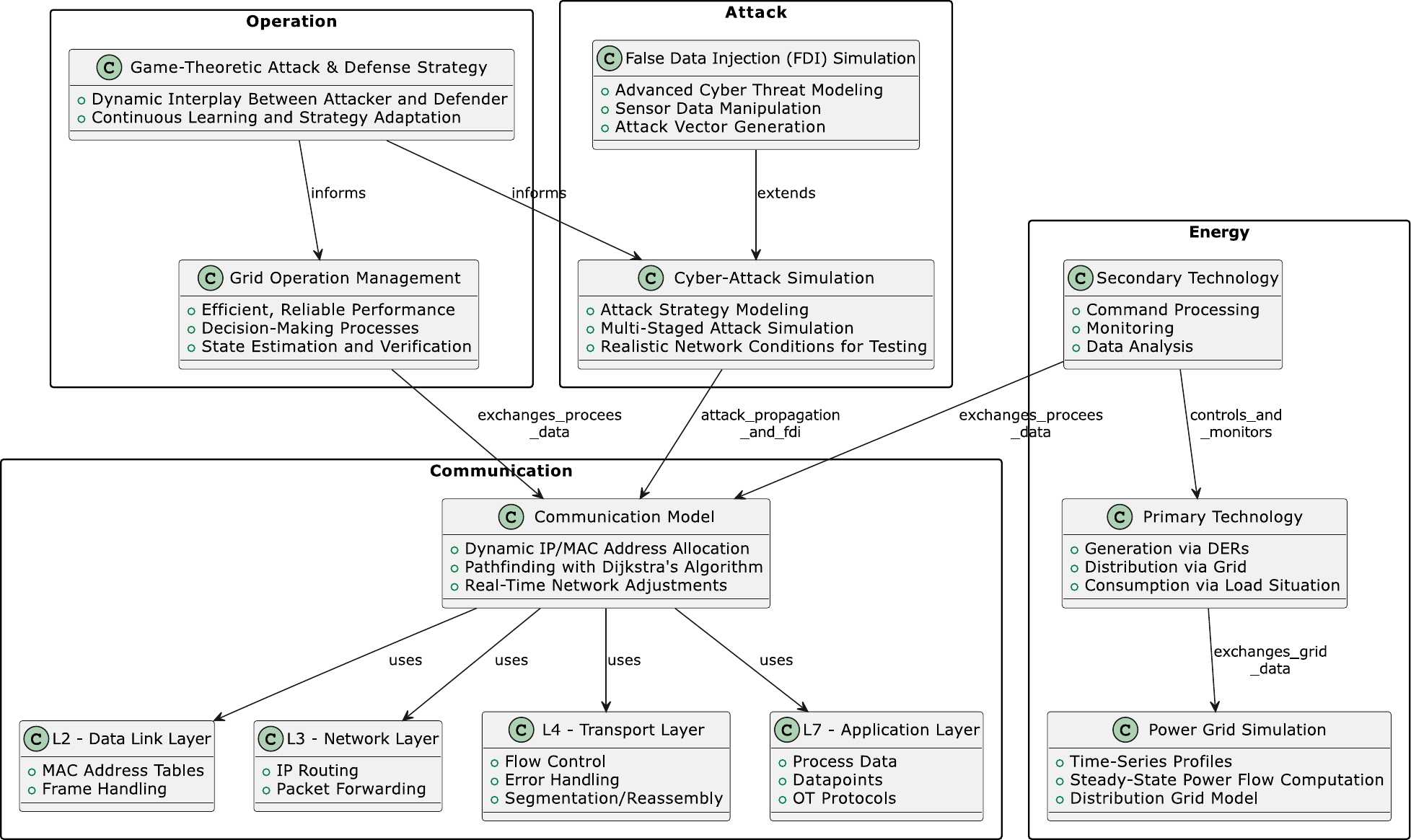}}
\caption{Comprehensive Overview of the Integrated Power Grid Communication and Simulation Model. This figure illustrates the multi-layered approach to power grid simulation, encompassing communication layers, operational strategies,  cyber security simulations, and energy grid technologies. }
\label{fig:sim_overview}
\end{figure*}

\section{Overview} \label{sec:overview}
Figure~\ref{fig:sim_overview} provides a comprehensive overview of the developed framework for understanding and simulating dynamic interactions between the cyber and physical planes within a power grid. It focuses on communication processes, operational strategies, cyber attack simulations, and the intrinsic functions of the power grid.

At the core of the model is the Communication Model, which handles crucial tasks such as dynamic IP/MAC address allocation, pathfinding with Dijkstra's algorithm \cite{dijkstra1959note}, and adjustments in real-time network settings. This model integrates with various network layers, including the Data Link Layer that manages MAC address tables and frame handling for network data forwarding; the Network Layer which handles IP routing and packet forwarding; the Transport Layer focuses on flow control, error handling, and data packet segmentation/reassembly; and the Application Layer that deals with process data, datapoint, and operational technology protocols.

The model also includes Grid Operation Management for ensuring efficient and reliable grid performance, covering decision-making processes, state estimation, and verification. A Game-Theoretic Attack \& Defense Strategy is represented, highlighting the complex interplay between attackers and defenders, with a focus on continuous learning and strategic adaptation. Additionally, the  Cyber Attack Simulation models realistic multi-staged attack strategies, examining network vulnerabilities and resilience, while the \ac{FDI} Simulation focuses on cyber threat modeling, sensor data manipulation, and attack vector generation.

Key elements of the framework also encompass Secondary Technology, which includes command processing, monitoring, and data analysis critical in controlling grid operations; Primary Technology involving generation, distribution, and consumption within the grid; and the Power Grid Simulation, which adds time-series profiles, steady-state power flow computation, and a detailed distribution grid model.

The framework features a network of interconnections, indicating the flow of information and command across various components. The Communication Model integrates with all network layers for data exchange, and it is utilized by both the  cyber attack Simulation and \ac{FDI} Simulation for simulating attacks. Grid Operation Management and the Game-Theoretic Attack \& Defense Strategy exchange data via the Communication Model. The Secondary Technology controls and monitors the Primary Technology and also exchanges data via the Communication Model. The Primary Technology interacts with the Power Grid Simulation for grid data exchange.

%% file: chapter2a.tex
\section{power grid Simulation} \label{sec:gridsimulation}
The Power Grid and Communication Simulation presents a novel, fully integrated approach that encapsulates the entire chain of power grid operation within a unified environment. This system models every stage, from data modeling, scenario generation, power flow simulation, to datapoint mapping and communication exchange, without relying on external devices or components, ensuring complete independence and internal consistency. The environment incorporates layer-based simulation grounded in the SGAM framework~\cite{klaer2020graph}, enabling the simulation of grid operation across physical (Energy Technology), communication (Information Technology), and operational (Operational Technology) layers, using PandaPower for power flow modeling and IEC 60870-5-104 over TCP/IP for communication.

What sets this system apart is its ability to simulate both the physical behavior of the grid and its communication dynamics within the same environment. The power flow simulation ensures accurate, real-time modeling of electrical grid operations in steady state, while the communication simulation mirrors real-world network conditions, handling protocols such as Ethernet, IP, and TCP. The communication layers, including \acp{RTU}, \ac{IED}, and switches, interact seamlessly with the physical grid, reflecting the data exchanges critical for grid operation. This cohesive simulation environment provides a comprehensive view of the smart grid's performance under normal conditions, ensuring accurate data flow and network management, all within one independent, closed-loop system.

In the following, we outline the methodology and approach within our paper used in the grid modeling and communication simulation.

\subsection{Grid Modelling} \label{subsec:gridsimulation_cps}
This section delves into the concepts of layer-based simulation environments and their significance in the overall grid simulation process. The foundation of this approach is grounded in our previous work on \ac{SGAM}-based graph modeling~\cite{klaer2020graph} and is completely implemented in Python.

\begin{figure*}\
\centerline{\includegraphics[width=\linewidth]{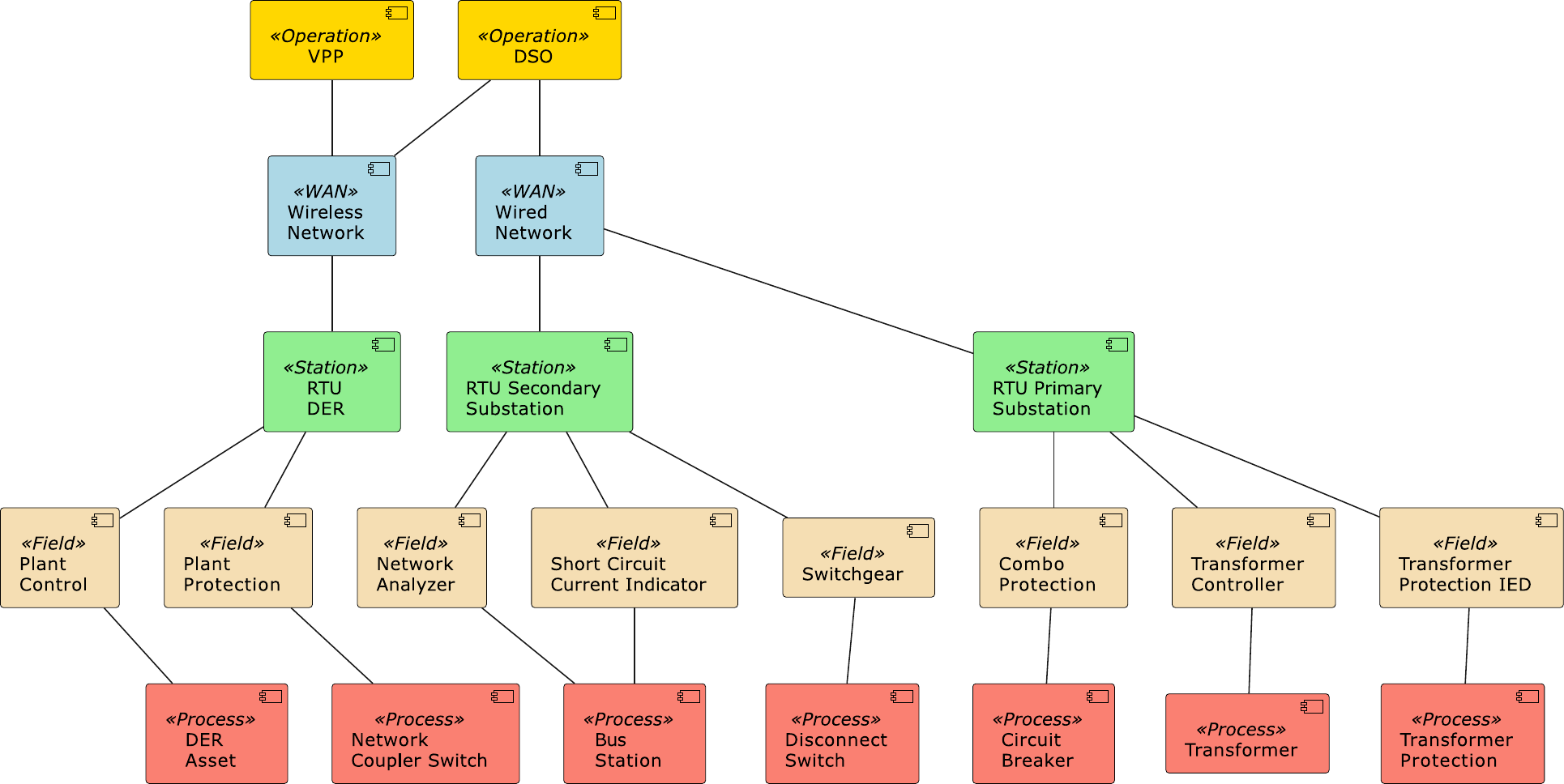}}
\caption{An exemplary illustration of the hierarchically object-orientated \ac{SGAM}-based modeling concept for grid simulation.}
\label{fig:sgam_model}
\end{figure*}

\paragraph{Concept of Layer-Based Simulation Environment}
The simulation environment's foundation is the \ac{SGAM}~\cite{uslar2019applying}, which systematically deconstructs the smart grid, allowing for the classification of system actors according to \ac{SGAM}'s three dimensions: Interoperability, Zones, and Domains. In this model, interoperability layers are categorized into \ac{ET}, \ac{IT}, and \ac{OT} domains (cf. Figure~\ref{fig:sgam_model}):

\begin{itemize}
    \item \ac{ET}: This encompasses the components and process levels, also known as Primary Technology.
    \item \ac{IT}: Represents the communication level, akin to Secondary Technology.
    \item \ac{OT}: Involves the information and function levels.
\end{itemize}

In \ac{SGAM}-based models, an attacker's influence spans across these domain-level layers, compromising devices in the \ac{IT} layer and manipulating the \ac{OT} layer to attack the \ac{ET} layer. The goal is to ensure that traces of attack propagation and execution are left across all levels of the interoperability axis.

The simulation environment must incorporate actors capable of creating, processing, and exchanging information. These actors should be classified into different domains and zones, reflecting the hierarchical organization of the system. Additionally, the introduction of an \ac{ICT} layer is essential to accurately simulate the \ac{WAN}, ensuring that communication dynamics are realistically represented.

\paragraph{Interoperability between Layers}
The environment uses a hierarchical communication structure, segregating the energy grid into distinct layers: Operation, \ac{WAN}, Station, Field, and Process. Each layer has its unique set of components and communication requirements, influencing the modeling approach.

The \ac{WAN}, bridging the Operation and Station layers, is characterized by a mix of wireless and wired communication networks. This duality necessitates a flexible modeling approach, accommodating the varying speeds, capacities, and reliability factors inherent in these communication modes. The \ac{DSO}'s interaction with both wireless and wired networks and the \ac{VPP}'s reliance on wireless communication, highlight the need for diverse communication protocols and robust network simulations.

At the Station layer, \acp{RTU} at primary and secondary substations, along with \ac{DER}, form crucial communication nodes. These \acp{RTU} serve as intermediaries, bridging the higher-level operational commands with the field-level implementations. Here, the communication network must efficiently model the transmission of operational commands and feedback between these layers. The distinct nature of each \ac{RTU}---whether it is part of a primary or secondary substation or integrated with \ac{DER}---requires tailored communication protocols and latency models.

The Field layer includes various protective and control devices such as transformer protection \acp{IED}, switchgear, and network analyzers. These devices communicate primarily with the Station layer, specifically the \acp{RTU}, to execute and report operational processes. Modeling this communication involves understanding the physical layout of the grid and the logical flow of information, which may vary depending on the device and its function within the grid.

At the base of the hierarchy lies the Process layer, encompassing physical grid elements such as transformers, circuit breakers, and disconnect switches. The communication here is more control-oriented, focusing on the direct management of physical assets. This layer's interaction with the Field layer devices dictates a need for precise, real-time communication modeling to ensure grid stability and responsiveness to operational changes.

This concept involves accurately modeling the inter-layer communications, from high-level operational decisions down to field-level device controls. The communication network must simulate various technologies, from wireless networks to traditional wired systems, ensuring comprehensive coverage of the grid's diverse communication needs. By doing so, the simulation can provide valuable insights into the grid's behavior under normal and stressed conditions, including potential cyber attack scenarios.

\paragraph{Layer Segregation} 
Our simulation environment leverages the NetworkX~\cite{hagberg2020networkx} library to construct a layered graph structure. It starts at the process level, represented by a PandaPower~\cite{thurner2018pandapower} network. The simulation uses a versatile PandaPower network to represent various electrical grid models, customizable in nodes, voltage levels, and substations for diverse analyses.

The hierarchical construction of the simulation environment begins with the generation of a NetworkX MultiGraph from the PandaPower network, serving as the process layer. This base layer then supports the construction of subsequent layers: The field layer, where \ac{IED} instances are assigned to nodes from the process layer; the station layer, grouping \ac{IED} nodes and associating them with corresponding \ac{RTU} nodes; and the \ac{WAN} layer, where Switch class instances are created and meshed based on the substation edges from the process layer.

The communication infrastructure's \ac{WAN} layer involves the integration of \ac{IT} devices such as routers, switches, and firewalls. 
This layer simulates \ac{L2} to \ac{L4} behaviors, corresponding to Ethernet, IP, and TCP protocols, respectively. 
Functionalities include routing and switching of TCP/IP packets between the operation and station layers.

Routing tables for communicable devices are created based on the shortest paths determined by the Dijkstra algorithm \cite{dijkstra1959note}. 
This algorithm identifies the paths to various endpoints (\ac{RTU} and \ac{MTU} instances) across the network, and these paths are stored as IP addresses in the routing tables.

\subsection{Communication Simulation} \label{subsec:gridsimulation_com}
Effective communication within the power grid is vital for its operational integrity, especially in the face of cyber threats. This section describes the simulation of communication protocols and the dynamics of data flow within the simulation environment.

\paragraph{Modeling of IT and OT Devices in Power Grids}
The simulation environment integrates a polymorphic approach for device representation, leveraging a class hierarchy that differentiates devices based on their operational layers and functionalities. The foundational class, Network Switches, is responsible for \ac{L2} functionalities, such as forwarding packets to their appropriate destinations using switching algorithms and rulesets to manage network traffic efficiently.

Routers, which inherit from the Network Switches class, operate at \ac{L3} by routing packets across different network segments. They examine incoming packets' IP addresses and determine the best path for transmission, either using the ARP table or the device's own routing table. The \ac{RTU} class, an extension of the Router class, introduces the ability to process packets at the application layer, encompassing command processing, monitoring data, and managing communication for operational control.

Dynamic IP and MAC address allocation are managed by both Router and Switch classes to facilitate efficient packet routing within the network's topology. The \ac{RTU} class plays a crucial role in dynamic address handling, assigning them based on network configuration and operational needs. This contributes to the precise and effective routing of packets in the network.

The environment uses algorithms such as Dijkstra's for pathfinding, ensuring data packets are transmitted through the most optimal routes based on current network conditions and the NetworkX Graph model's topology. At \ac{L2}, switches use MAC addresses for frame forwarding and manage Source-Address-Tables autonomously, routing packets accurately within the network.

\acp{RTU}, within the application layer, are modeled to handle various tasks such as command processing, monitoring, and data analysis. As critical nodes, they facilitate the receiving and sending of control commands and data, vital for the power grid's operations. Their interactions with other network elements mirror realistic data exchange processes in the grid.

The environment enables comprehensive modeling of command transmissions and monitoring processes, integral for assessing the impact of  cyber attacks on power grids. These processes are implemented using event-driven methods within the \ac{RTU} class, reflecting the interactions across different communication model layers. Moreover, the environment is designed for real-time adaptability, adjusting routing paths in response to changing network conditions such as congestion or cyber attacks, thus providing insights into the power grid's resilience under various scenarios.

Network congestion simulations and subsequent data flow and routing adjustments are also effectively modeled. These include modifying routing tables, employing backup paths, and utilizing congestion control protocols, ensuring the network's continuous and efficient communication even under strained conditions.

The simulation employs key communication protocols such as Ethernet for \ac{L2}, IP for \ac{L3}, TCP for \ac{L4}, and IEC 60870-5-104 for \ac{L7}, modeled to mimic real-world network conditions, offering a realistic testing and analysis environment.

\paragraph{Simulation of Communication Protocols}
\begin{figure}\
\centerline{\includegraphics[width=\linewidth]{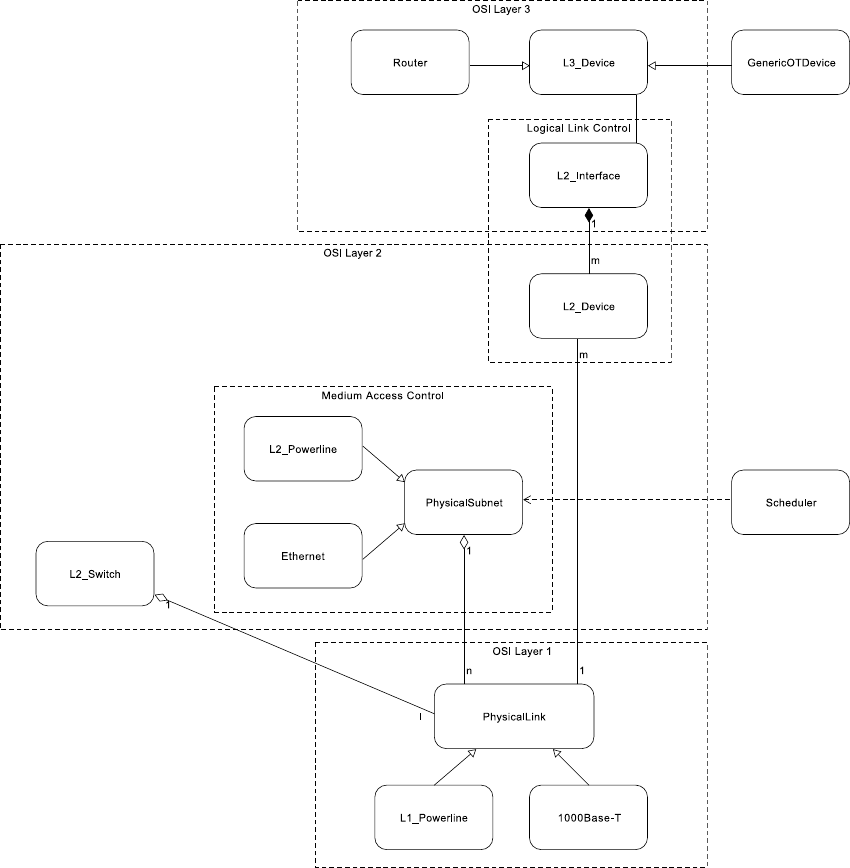}}
\caption{Overview of the communication model within the simulation environment. It successfully integrates various aspects of networking, from routing and switching to application layer processes. }
\label{fig:comm_model}
\end{figure}

\begin{table}[ht]
\centering
\caption{Description of Network Layers and Protocols}
\label{tab:network_layers}
\begin{tabular}{|p{1cm}|p{2cm}|p{4cm}|}
\hline
\textbf{Layer} & \textbf{Reference} & \textbf{Description} \\ \hline
CSMA/ CA /CD & \cite{LUN20} & Carrier Sense Multiple Access protocols used for controlling media access in network environments. \\ \hline
Ethernet & IEEE8023\cite{IEEE8023} & A widely used technology in local area networks that manages packet transmission over different network segments. \\ \hline
WLAN & IEEE80211\cite{IEEE80211} & Wireless Local Area Network protocol, enabling wireless connectivity in local areas. \\ \hline
MAC & IEEE8024\cite{IEEE8024} & Media Access Control, part of the data link layer that manages protocol access to the physical network medium. \\ \hline
ARP & RFC826\cite{RFC826} & Address Resolution Protocol used for mapping a network address (IP address) to a physical address (MAC address). \\ \hline
IPv4 & RFC791\cite{RFC791} & Internet Protocol version 4, used for addressing and routing data across networks. \\ \hline
ICMP & RFC792\cite{RFC792} & Internet Control Message Protocol used for error reporting and query messages in network communication. \\ \hline
TCP & RFC793\cite{RFC793} & Transmission Control Protocol, providing reliable, ordered, and error-checked delivery of data between applications. \\ \hline
UDP & RFC768\cite{RFC768} & User Datagram Protocol, used for simple transmission of datagrams without acknowledgments or guaranteed delivery. \\ \hline
\end{tabular}
\end{table}

Within the simulation environment, \ac{L2} to \ac{L4} are modelled according to the illustration in Figure~\ref{fig:comm_model} and Table~\ref{tab:network_layers}.
In the integrated simulation environment, Ethernet switches are modeled to reflect their function in real-world networks. This includes the management of MAC tables, essential for addressing and directing data packets to the correct destination. The simulation also contemplates the handling of frames and the concept of collision domains, highlighting the switches' role in controlling data flow within \ac{L2} of the OSI model. This modeling approach is critical for studying the impact of  cyber attacks on power grids, providing a realistic framework for network behavior analysis.

The IP routing mechanism in the simulation is designed to showcase both static and dynamic routing processes. Using the PandaPower and NetworkX Graph models, the simulation dynamically generates network participants based on predefined rules, encompassing both \ac{L2} and \ac{L3} devices. This approach allows for an elaborate examination of IP routing tables' management, providing insights into how packets are directed across various network segments. The simulation's focus on routing mechanisms aids in comprehending how  cyber attacks could potentially disrupt network traffic in power grids.

The simulation extends to higher layers of the OSI model, \ac{L7}, particularly focusing on application protocols (e.g. IEC 60870-5-104) vital for power grid operations. This includes a thorough simulation of the data exchange between the \ac{OT} device classes and other network elements, implemented through event-driven methods. The framework allows for transparent processes and interfaces, enabling the simulation of cyber attacks on lower protocol layers. By simulating different communication technologies and their specific vulnerabilities, the environment provides a comprehensive tool for analyzing the resilience of power grids against cyber threats.

Furthermore, the simulation includes the implementation of UDP over the typical TCP protocol, reducing complexity and maintaining compatibility with the rest of the simulation environment. The \ac{L3} devices within the simulation are programmed to send and receive packets using generic methods that interact directly with the assigned \ac{L2} interface, demonstrating a detailed interplay between the various layers of network communication. This layered approach in the simulation offers a multifaceted perspective on how different protocols and services operate within a power grid infrastructure.

%% file: chapter2b.tex
\section{Cyber Security Simulation} \label{sec:cssim}
\begin{table}[ht]
\centering
\caption{Nomenclature}
\begin{tabular}{p{1.7cm}p{6.5cm}}
\hline
Symbol & Description \\
\hline
$W_{i,j}$ & Weight of the edge connecting nodes $i$ and $j$ \\
$t$ & Time needed for a particular step \\
$C_{j}^{attacker}$ & Outage costs for the component from the attacker's viewpoint \\
$P_{j}^{attacker}$ & Likelihood of successfully compromising a node \\
$Risk$ & Risk of grid operation disruption due to cyber attack \\
$Q_{i}$ & Learning rate for each node $i$ \\
$c_{CB}(v)$ & Current flow betweenness centrality \\
$\tau_{st}(v)$ & Throughput from node $s$ to node $t$ via node $v$ \\
$b_{st}(v)$ & Absolute value of the total amount of current that flows through $v$ \\
$r(\vec{e_{i, j}})$ & Resistance between nodes $i$ and $j$ \\
$c_i ^ {outage}$, $c_j ^ {outage}$ & Outage costs of nodes $i$ and $j$ \\
$TTC(s, W)$ & Time to Compromise \\
$t_1$, $t_2$ & Time taken for the first and second stages of an attack \\
$P_1$ & Probability of the first stage of an attack \\
$u$ & Unsuccessful rate of the second stage of an attack \\
$N$ & Total number of nodes \\
$V_i$ & Node voltage \\
$V_{nominal}$ & Nominal voltage \\
$\Delta V_{max}$ & Maximum allowable voltage deviation \\
$I_{line}$ & Current in the line \\
$I_{max}$ & Maximum allowable current \\
$\hat{z}$ & Estimated measurement vector \\
$H$ & Jacobian matrix \\
$x$ & State vector \\
$P_{i}^{min}$, $P_{i}^{max}$ & Minimum and maximum active power outputs of generator $i$ \\
$Q_{i}^{min}$, $Q_{i}^{max}$ & Minimum and maximum reactive power outputs of generator $i$ \\
$ci$ & Cost coefficient for generator $i$ \\
$\vec{a}$ & Attack vector \\
$\vec{c}$ & Non-zero vector representing deviation from estimated normal state \\
$\hat{x}$ & True state \\
$\hat{x}_a$ & Estimated state with manipulation \\
$r$ & Residual without manipulation \\
$r_a$ & Residual with manipulation \\
$P$ & Set of protected measurements \\
$\alpha_i$ & Minimization criteria for the attack vector \\
\hline
\end{tabular}
\label{tab:nomenclature}
\end{table}

Cybersecurity simulation is crucial for comprehending the vulnerabilities and threats within power grid operations. This section outlines the methodologies for simulating grid operations and multi-staged cyber attacks, encompassing both \ac{FDI} and \ac{OT}-related attacks. The multi-stage attack represents the initial phase of the attacker, characterized by the propagation and spreading of influence, whereas the \ac{FDI} constitutes the final stage, involving coordinated attacks and the execution of technically adverse impacts on the grid~\cite{musleh2019survey}. Table~\ref{tab:nomenclature}  presents the nomenclature of the variables used in this work.

\subsection{Grid Operation Simulation} \label{subsec:cssim_op}
The management of electric energy distribution through power grids involves a complex array of operations, technologies, and methodologies to ensure efficient, reliable grid performance. These networks typically comprise radial ring circuits operated in an open ring format under normal conditions. Adhering to the (n-1) criteria, they maintain grid security and reliability, allowing for temporary overloads up to 130\% in case of operational disturbances:
\begin{equation}
    |V_i - V_{nominal}| \leq \Delta V_{max}
\end{equation}
\begin{equation}
        I_{line} \leq I_{max}
\end{equation}
Where $V_i$ is the node voltage, $V_{nominal}$ is the nominal voltage, $\Delta V_{max}$  is the maximum allowable voltage deviation, $I_{line}$  is the current in the line, and $I_{max}$  is the maximum allowable current.
  
Operational management in grid systems demands continual monitoring and adjustments of grid states to uphold operational intervals such as voltage bands and thermal limits. This management is particularly crucial during disturbances, where rapid response measures such as topology changes, transformer tap adjustments, and reactive power compensation are enacted. These include adapting grid configuration to redistribute loads, modifying transformer voltage levels, and utilizing connected compensation equipment for voltage regulation.

\begin{figure*}\
\centerline{\includegraphics[width=\linewidth]{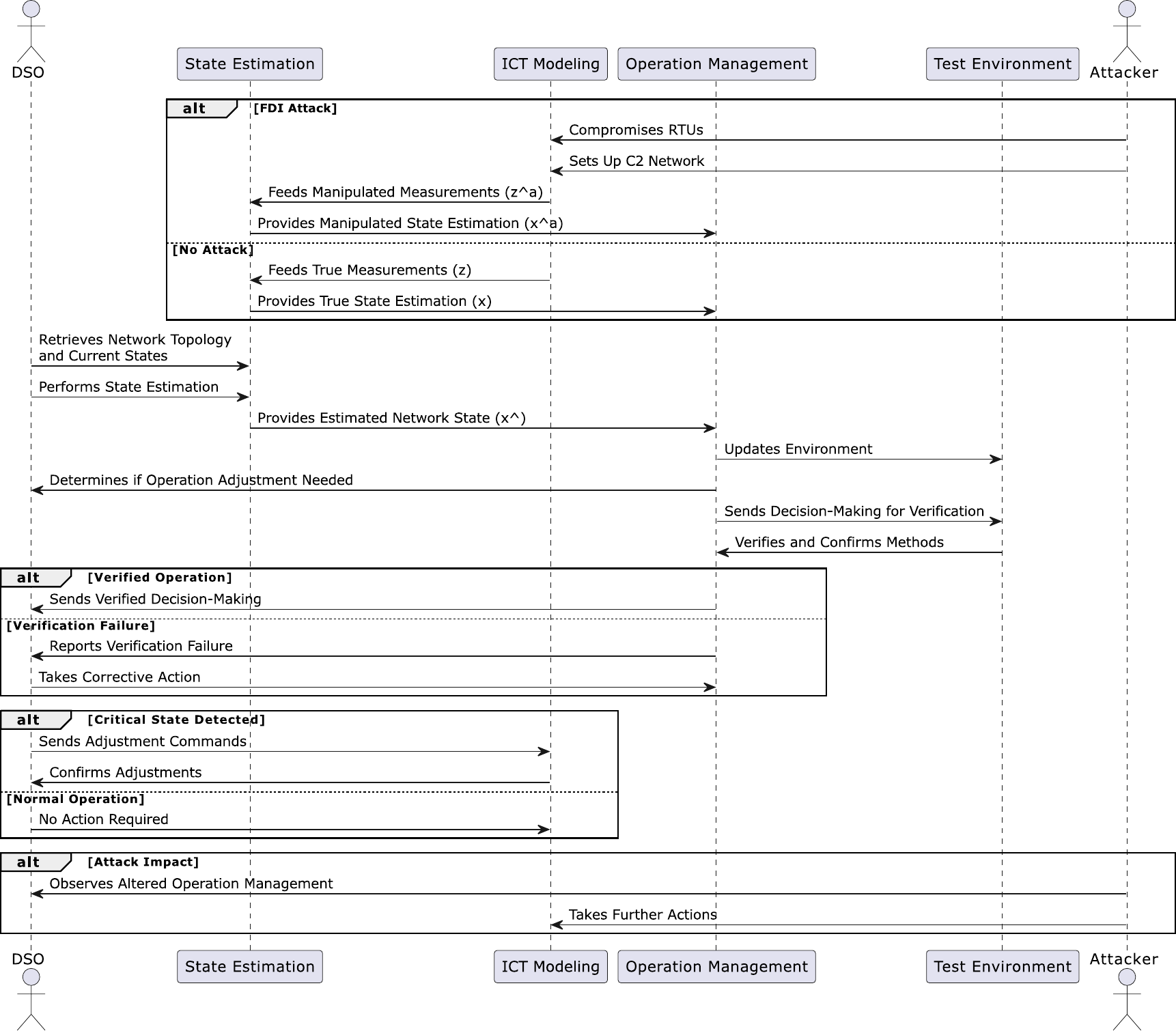}}
\caption{Sequence diagram illustrating grid operation management with verification process in response to \ac{FDI} attacks and normal conditions. }
\label{fig:grid_op_seq}
\end{figure*}

\ac{SE} is integral to grid operation management~\cite{alam2014distribution}, providing accurate grid state snapshots using real-time, pseudo, and virtual measurements. Here, $\hat{z}$ is the estimated measurement vector, $H$ is the Jacobian matrix, and $x$ is the state vector.
\begin{equation}
    \hat{z} = Hx
\end{equation}
While traditionally used in transmission grids, \ac{SE} is increasingly applied in distribution grids due to power grid advancements. High-level decision-making processes such as short circuit current computations, grid \ac{SE}, topology adjustments, and voltage-reactive power optimization are essential for economic optimization, availability, and safety in grid management.
Figure~\ref{fig:grid_op_seq} illustrates the essence of the sequence diagram, highlighting the process of grid operation management and the inclusion of a verification step, as well as how the system responds to both \ac{FDI} attacks and standard operational conditions.

Grid operation encompasses overseeing and controlling the grid, identifying disturbances, alleviating congestion, and planning maintenance and revisions. The grid operator ensures compliance with defined parameters such as voltage limits and thermal capacities of components. Methods such as load flow calculations and \ac{OPF} are employed for efficient energy distribution and cost minimization. Here, $c_i$ is the cost coefficient for generator $i$ and $P_i$ and $Q_i$ are the active and reactive power outputs of generator $i$.
\begin{equation}
    \min \sum_{i \in G} c_i P_i^2
\end{equation}
\begin{equation}
    \text{s.t. } P_{i, min} \leq P_i \leq P_{i, max}
\end{equation}
\begin{equation}
    \text{s.t. } Q_{i, min} \leq Q_i \leq Q_{i, max}
\end{equation}
\ac{SE} aids in verifying these calculations by providing a more accurate grid state derived from various measurements. In \ac{SE}, \ac{BDD} algorithms are crucial for identifying and correcting erroneous measurements, ensuring the reliability of \ac{SE} results. \ac{BDD}algorithms use residual-based detectors to compare observed measurements against estimated values, detecting significant discrepancies.

The modeling of grid operation management particularly emphasizes medium-voltage networks, focusing on the role of the \ac{DSO}. It assumes the \ac{DSO} has comprehensive knowledge of the grid state (topology and state variables), essential for operational decision-making. Operational constraints are defined, such as allowable voltage deviations and maximum line thermal capacities, ensuring a safe operation. The \ac{DSO} employs specific algorithms for topology changes (cf. Algorithm~\ref{alg:loadred}) and transformer tap adjustments (cf. Algorithm~\ref{alg:tapchanger}), adapting the grid to maintain operational norms and respond effectively to disturbances~\cite{haque2017smart, schneider2020analysis}. 

\begin{algorithm}
\caption{Simulation of Load Redistribution for Topology Changes}
\label{alg:loadred}
\begin{algorithmic}[1]
    \State Close all switches in the power network
    \State Perform a load flow calculation
    \State Determine a minimum spanning tree using Kruskal's algorithm
    \State Select the line with the least load from the minimum spanning tree
    \State Disconnect the selected line
    \State Perform another load flow calculation
    \State Check for compliance with the operational limits
\end{algorithmic}
\end{algorithm}

\begin{algorithm}
\caption{Determination of the Optimal Tap Changer Position}
\label{alg:tapchanger}
\begin{algorithmic}[1]
\For{$i = 1$ \textbf{to} \texttt{maxNumberOfTaps}}
    \State Identify node with voltage band violation
    \State Identify the nearest transformer
    \If{Overvoltage \textbf{and} tap position \textbf{not} maximal}
        \State tap position \texttt{+= 1}
    \ElsIf{Undervoltage \textbf{and} tap position \textbf{not} minimal}
        \State tap position \texttt{-= 1}
    \Else
        \State \textbf{continue}
    \EndIf
    \State Perform a load flow calculation
\EndFor
\end{algorithmic}
\end{algorithm}

The optimal reconfiguration of the grid involves calculated steps using algorithms such as Kruskal's for minimal spanning trees~\cite{greenberg1998greedy} to redistribute loads and alleviate overloads. The focus is also on decentralized generation management, where the \ac{DSO} regulates injections from \acp{DER}, balancing economic considerations with operational necessities.

\subsection{Multi-Staged Attack Simulation} \label{subsec:cssim_mulatt}
In the simulation environment's multi-stage attack simulation, Algorithms~\ref{alg:attack_sim} and \ref{alg:attack_propagation} illustrate the precise methodologies attackers use to infiltrate and destabilize critical systems. The abstract formalism is based on prior works~\cite{sen2023approach}.

The algorithms delineate the attacker model as it systematically executes operations to reach malicious goals. Modeled after the MITRE \ac{ICS} ATT\&CK Matrix~\cite{alexander2020mitre} and the SANS-Ukraine-Killchain~\cite{whitehead2017ukraine, capano2019understand}, the algorithms embedded in the simulation demonstrate the attacker's process for system detection, gradual network compromise, and coordinated exploitation based on identified system weaknesses.

The algorithms deploy the \ac{MulVAL} framework~\cite{b10} to generate logical attack paths that reflect the attacker's tactical navigation within the network. This process is seamlessly integrated with the simulation environment's implementation of the \ac{ICS} Kill-Chain stages. It orchestrates the stepwise sequence of exploitation, malware deployment, establishment of command and control, and execution of actions determined by the \ac{C2} node, as facilitated by the algorithms.

The algorithms abstract real-world attack strategies, such as those employed by Havex~\cite{venkatachary2017economic} and Stuxnet~\cite{baezner2017stuxnet}, fitting them within the simulation environment's digital confines (see prior works~\cite{sen2022using}). This maintains the authenticity of the attack techniques while adhering to the simulated constraints, ensuring a realistic depiction of the potential impact on virtual representations of physical grid systems.

Through the algorithms, a \ac{C2} node is established within the network to act as the hub of the attack, directing compromised hosts and centralizing intelligence gathering---a pivotal element in the attack narrative dictated by the algorithms.

The simulation environment's adaptability is underscored by the algorithms' capability to model a variety of attack strategies, including non-pattern-based attacks, simulating the erratic nature of real-world threats. These dynamic strategies challenge the system's defenses and evaluate the robustness of their security concepts.

At the heart of the simulation lies the iterative propagation algorithm, enabling the simulated attacker to modify its route in response to network defenses such as firewall configurations. This reflects a realistic approach to circumventing or overcoming security measures within the target network.

The algorithms culminate in a definitive outcome that either disrupts the system or achieves another specific aim. This event marks the transition to the next phase of the attack---typically the energy attacker phase---highlighting the layered nature of the threat. The results of the algorithms' execution offer detailed insights into the attack's impact, vital for enhancing detection and response strategies within the simulation environment.

Through these features, the simulation environment's algorithms offer a sophisticated and actionable portrayal of cyber threats, providing an essential resource for improving cyber security in critical infrastructure settings.

\begin{algorithm}
\caption{Initialization and Attack Simulation}
\label{alg:attack_sim}
\begin{algorithmic}[1]
\State \textbf{Global} C2Host, ICS\_KillChainStages, SCADASubnets, AttackerGoals
\State Initialize the SCADA environment with network subnets and attack scenarios
\State C2Host $\gets$ InitializeC2()
\State ICS\_KillChainStages $\gets$ SetKillChainStages()
\State AttackerGoals $\gets$ SetAttackerGoals()
\State SCADASubnets $\gets$ InitializeSubnets()
\State Attacker $\gets$ InitializeAttacker(C2Host)

\Procedure{Attack}{$SCADASubnets$, $C2Host$}
    \For{each subnet in SCADASubnets}
        \State PropagateAttack(subnet, Attacker)
    \EndFor
    \State EvaluateOutcome(AttackerGoals, SCADASubnets)
\EndProcedure

\Function{InitializeC2}{}
    \State C2 $\gets$ CreateNewC2Node()
    \State InstallC2Malware(C2)
    \State \Return C2
\EndFunction

\Function{SetKillChainStages}{}
    \State Stages $\gets$ DefineStagesBasedOnRealAttacks()
    \State \Return Stages
\EndFunction

\Function{SetAttackerGoals}{}
    \State Goals $\gets$ DefineGoalsBasedOnMITREandSANS()
    \State \Return Goals
\EndFunction

\Function{InitializeSubnets}{}
    \State Subnets $\gets$ CaptureSubnetDetailsFromSCADA()
    \State \Return Subnets
\EndFunction

\Function{EvaluateOutcome}{$Goals$, $Subnets$}
    \State Assessment $\gets$ EvaluateAgainstGoals(Goals, Subnets)
    \State \Return Assessment
\EndFunction
\end{algorithmic}
\end{algorithm}

\begin{algorithm}
\caption{Attack Propagation within Subnets}
\label{alg:attack_propagation}
\begin{algorithmic}[1]
\Procedure{PropagateAttack}{$Subnet$, $Attacker$}
    \State HostsToAttack $\gets$ IdentifyHosts(Subnet)
    \For{each host in HostsToAttack}
        \State ExploitVulnerabilities(host, Attacker)
        \State InstallMalware(host, Attacker)
        \State EstablishC2Communication(host, C2Host)
        \State ExecuteC2Commands(host, C2Host)
        \If{Primary target in subnet?}
            \State Generate attack graph
            \State Create attack trace
        \Else
            \While{Nodes connecting to other subnets}
                \If{Subnet fully explored?}
                    \If{Unexplored subnets known?}
                        \State Move to new subnet
                        \State Create new sub-goals
                    \Else
                        \State Mark interesting nodes in subnet
                    \EndIf
                \Else
                    \State Mark nodes
                    \State Mark nodes leading elsewhere
                \EndIf
            \EndWhile
            \If{No new sub-goals}
                \State Abort the attack propagation
            \EndIf
        \EndIf
        \State DetermineOutcome(Attacker)
    \EndFor
\EndProcedure

\Function{IdentifyHosts}{$Subnet$}
    \State Hosts $\gets$ DiscoverHostsInSubnet(Subnet)
    \State \Return Hosts
\EndFunction

\Function{ExploitVulnerabilities}{$Host$, $Attacker$}
    \State Vulnerabilities $\gets$ GetHostVulnerabilities(Host)
    \State Attacker.Exploit(Vulnerabilities)
\EndFunction

\Function{InstallMalware}{$Host$, $Attacker$}
    \State Attacker.Install(Host)
\EndFunction

\Function{EstablishC2Communication}{$Host$, $C2$}
    \State Attacker.Communicate(Host, C2)
\EndFunction

\Function{ExecuteC2Commands}{$Host$, $C2$}
    \State Commands $\gets$ C2.IssueCommands()
    \State Attacker.Execute(Host, Commands)
\EndFunction

\Function{DetermineOutcome}{$Attacker$}
    \State Outcome $\gets$ AnalyzeAttackEffectiveness(Attacker)
    \State \Return Outcome
\EndFunction
\end{algorithmic}
\end{algorithm}

%% file: chapter2c.tex
\subsection{False Data Injection Attack}\label{subsec:cssim_fdi}
In the simulation environment, the \acf{FDI} attack is methodically modeled and executed as an advanced cyber threat targeting the electrical energy information systems integrated within \ac{CPS}. The simulation incorporates a layered approach reflecting the \ac{CPS}'s structure, consisting of the perception execution layer, data transmission layer, and application control layer. These layers correspond to sensor/actuator networks, communication protocols, and operational control, respectively.

The \ac{FDI} attack simulation introduces the manipulation of sensor data, aiming to induce a controlled response from the system's control mechanisms (cf. Algorithm~\ref{algo:fdi}). This is achieved by meticulously generating an attack vector that alters specified measurements without triggering any \ac{BDD} alarms. The simulation adheres to the historical context of \ac{FDI} attacks, also known as stealthy deception attacks, load redistribution attacks, or data integrity attacks, to create a realistic and stealthy operational disruption.

An \ac{FDI} attack in the simulation requires the attacker to adjust a certain number of measurements to influence state variables successfully. The seminal work by~\cite{liu2011false} is harnessed to form an attack vector that bypasses \ac{BDD} alarms. The vector is constructed using a structured approach to avoid random anomalies, which would typically trigger alarms.

The simulation outlines the relationship between the actual state estimation with and without manipulated data, maintaining the same $\| \cdot \|_2$ norm of the residual, ensuring no \ac{BDD} alarms are activated. This aspect of the simulation emphasizes the difficulty in detecting such attacks due to their low observability.

To execute an \ac{FDI} attack without triggering alarms, the simulation devises a structured attack vector following the equations \ref{eq:avec}-\ref{eq:bddsup}. The attack vector formulation in an \ac{FDI} attack is a crucial step where a vector $\vec{a}$ is created to manipulate specific measurements. This is done through $\vec{a} = H \cdot \vec{c}$, where $H$ is the Jacobian matrix, and $\vec{c}$ is a non-zero vector representing the deviation from the estimated normal state. The state estimation with manipulated data involves recalculating the estimated state $\hat{x}_a = \hat{x} + \vec{c}$, where $\hat{x}$ is the true state, to reflect the impact of the attack. The $\| \cdot \|_2$ Norm of the residual with manipulated measurements, $|r_a|_2$, ensures that the manipulated measurements do not trigger \ac{BDD} alarms, maintaining $|r_a|_2 = |r|_2$, where $r$ is the residual without manipulation. The definition of the attack vector as per the manipulation target is expressed by setting $a_i = H[i] \cdot \vec{c} = 1$ for the i-th targeted measurement, aligning the attack focus. This is subject to the constraint that the alteration of the i-th measurement must match the desired outcome, while the definition of the \ac{BDD} alarms suppression condition ensures that $a_k = H[k] \cdot \vec{c} = 0$ for all $k \in P$, where $P$ is the set of protected measurements, to avoid detection.

Let $H$ be the Jacobian matrix, $\vec{a}$ be the attack vector, $\vec{c}$ be a non-zero vector, $\hat{x}$ be the true state, $\hat{x}_a$ be the estimated state with manipulation, $r$ be the residual without manipulation, and $r_a$ be the residual with manipulation. Let $P$ denote the set of protected measurements.

\begin{equation}
\label{eq:avec}
\vec{a} = H \cdot \vec{c}
\end{equation}

\begin{equation}
\hat{x}_a = \hat{x} + \vec{c}
\end{equation}

\begin{equation}
\begin{aligned}
    \| r_a \|_2 = \| z_a - H \hat{x}_a \|_2 = \\
    \| (z + \vec{a}) - H(\hat{x} + \vec{c}) \|_2 =  \\
    \| z - H\hat{x} \|_2 = \| r \|_2
\end{aligned}
\end{equation}

\begin{equation}
\alpha_i = \min \| \vec{a} \|_0 = \min_{\vec{c}} \| H \cdot \vec{c} \|_0
\end{equation}

\begin{equation}
\text{s.t. } a_i = H[i] \cdot \vec{c} = 1
\end{equation}

\begin{equation}
\label{eq:bddsup}
\text{s.t. } a_k = H[k] \cdot \vec{c} = 0 \quad \forall k \in P
\end{equation}

The simulation then navigates through the attacker's goals, using deception tactics to simulate critical conditions such as an overloaded transmission line, or concealment attacks to mask actual critical states. The objective is to make the system operators believe that the grid is functioning normally or in a critical state, depending on the attack's intention.

The attacker model within the simulation environment operates under precise conditions. It leverages knowledge of the grid topology and accessible measurements to craft the minimum attack vector that influences the desired state variable without raising suspicion. This formulation transforms into an optimization problem, where the attacker seeks to minimize the non-zero elements of the attack vector while ensuring the desired alteration of a specific measurement.

Finally, the simulation investigates the operational processes within the \ac{CPS}. It scrutinizes whether manipulated measurements can cause changes in grid operational procedures and if such altered actions could plunge the actual energy grid into a critical state, revealing the severity of \ac{FDI} attacks' impact on the operation and safety of \ac{CPS}.

\begin{algorithm}
\caption{FDI Attack Algorithm}
\label{algo:fdi}
\begin{algorithmic}[1]
\State \textbf{Input:} Network Topology, Sensor Data $z$, Jacobian Matrix $H$, Protected Set $P$
\State \textbf{Output:} Manipulated Measurements $z_a$

\Procedure{FDIVector}{$H$, $TargetMeasurement$, $P$}
    \State Initialize $\vec{c}$ as a non-zero vector
    \State Formulate the attack vector $\vec{a} = H \cdot \vec{c}$
    \State Set $a_i = H[i] \cdot \vec{c} = 1$ for $i = TargetMeasurement$
    \ForAll{$k \in P$}
        \State Ensure $a_k = H[k] \cdot \vec{c} = 0$ to suppress BDD alarms
    \EndFor
    \State Minimize the number of non-zero entries in $\vec{a}$, $\min \| \vec{a} \|_0$
    \State \Return $\vec{a}$
\EndProcedure

\Procedure{FDI}{$z$, $H$, $TargetMeasurement$, $P$}
    \State $\vec{a} \gets$ \Call{FDIVector}{$H$, $TargetMeasurement$, $P$}
    \State Create manipulated measurements $z_a = z + \vec{a}$
    \State Estimate manipulated state $\hat{x}_a = \hat{x} + \vec{c}$ where $\hat{x}$ is the true state
    \State \Return $z_a$
\EndProcedure

\Procedure{FDIAttackSimulation}{}
    \State Load grid topology and sensor data $z$
    \State Compute Jacobian matrix $H$ for the grid
    \State Define protected set $P$
    \State Choose a target measurement for manipulation
    \State $z_a \gets$ \Call{FDI}{$z$, $H$, $TargetMeasurement$, $P$}
    \State Apply $z_a$ to the grid state estimation process
    \State Analyze and compare the results with the true state
\EndProcedure
\end{algorithmic}
\end{algorithm}

\subsection{OT-Related Attacks}\label{subsec:cssim_ot}
Along with the \ac{FDI} attack, the \ac{OT}-related attack pertains to the operational aspect, directly impacting the grid. Concurrently, the \ac{FDI} conceals these effects from the operator's perspective.
In the \ac{OT}-level attack modeling, a central coordinator is employed. This coordinator determines actions for devices that compromise the grid based on information from devices compromised in the \ac{IT} attack propagation phase.

The grid's state is evaluated using the classification, which defines four main classes based on given threshold gradations and the grid's maximum and minimum size positions. These include:
\begin{itemize}
    \item Class 0: Bus voltages and operational equipment utilization are within the first threshold settings, and there are no unserved loads in the grid.
    \item Class 1: Bus voltages or operational equipment utilization violate the first threshold settings but remain within the second. No unserved loads exist.
    \item Class 2: Second threshold settings are violated for bus voltages or equipment utilization. Still, no unserved loads exist.
    \item Class 3: Unserved loads are present.
\end{itemize}

Assuming the coordinator is aware of the complete physical grid, its state, and topology, it determines iterative bad-case scenarios. The extent of these scenarios is defined by the attacker's operational range within the grid, which depends on the location and number of compromised devices.

The coordinator prioritizes the types of compromised devices based on the hierarchy in power grids. Devices from the field level are prioritized lower than those from the station level, and so on. Priority is used to avoid overlaps in responsible grid areas. Actions are determined only for the highest priority devices that can influence grid areas not covered by other devices.

The coordinator executes changes via the Pandapower grid lines, influenced by the compromised devices. These changes include:
\begin{itemize}
    \item Opening circuit breakers.
    \item Changing the output of \acp{DER} to arbitrary values between 0\% and 100\%.
    \item Changing the $\cos\phi$ of \acp{DER} for reactive power adjustment or voltage change.
    \item Altering transformer tap settings based on transformer voltage.
\end{itemize}

Based on a snapshot of the Pandapower grid, specified changes are made. The grid's state class is checked to be worse than the initial. If correct, changes are stored as command actions in a collection. This collection comprises the commands that would have the most severe impact on the grid state.

If the coordinator lacks sufficiently influential compromised devices, false monitoring, and measurement actions are created based on the physical grid state. This involves creating an altered grid image via exceeding threshold values or applying previously described command actions.

%% file: chapter3.tex
\section{Strategies for Cyber Attack and Defense} \label{sec:framework}
In this study, we introduce a novel methodology for the generation of synthetic data related to cyber attacks. This methodology employs a game-theoretic framework to analyze interactions between attackers and defenders. In particular, we model the dynamics between an attacker and defender in a power grid intrusion scenario, employing key terms such as ``starting capital'', ``funds'', ``betweenness centrality'', and ``path optimization'' to analyze their capabilities and strategic decision-making.

\subsection{Conceptual Overview} \label{subsec:framework_overview}
\begin{figure*}
\centerline{\includegraphics[width=\linewidth]{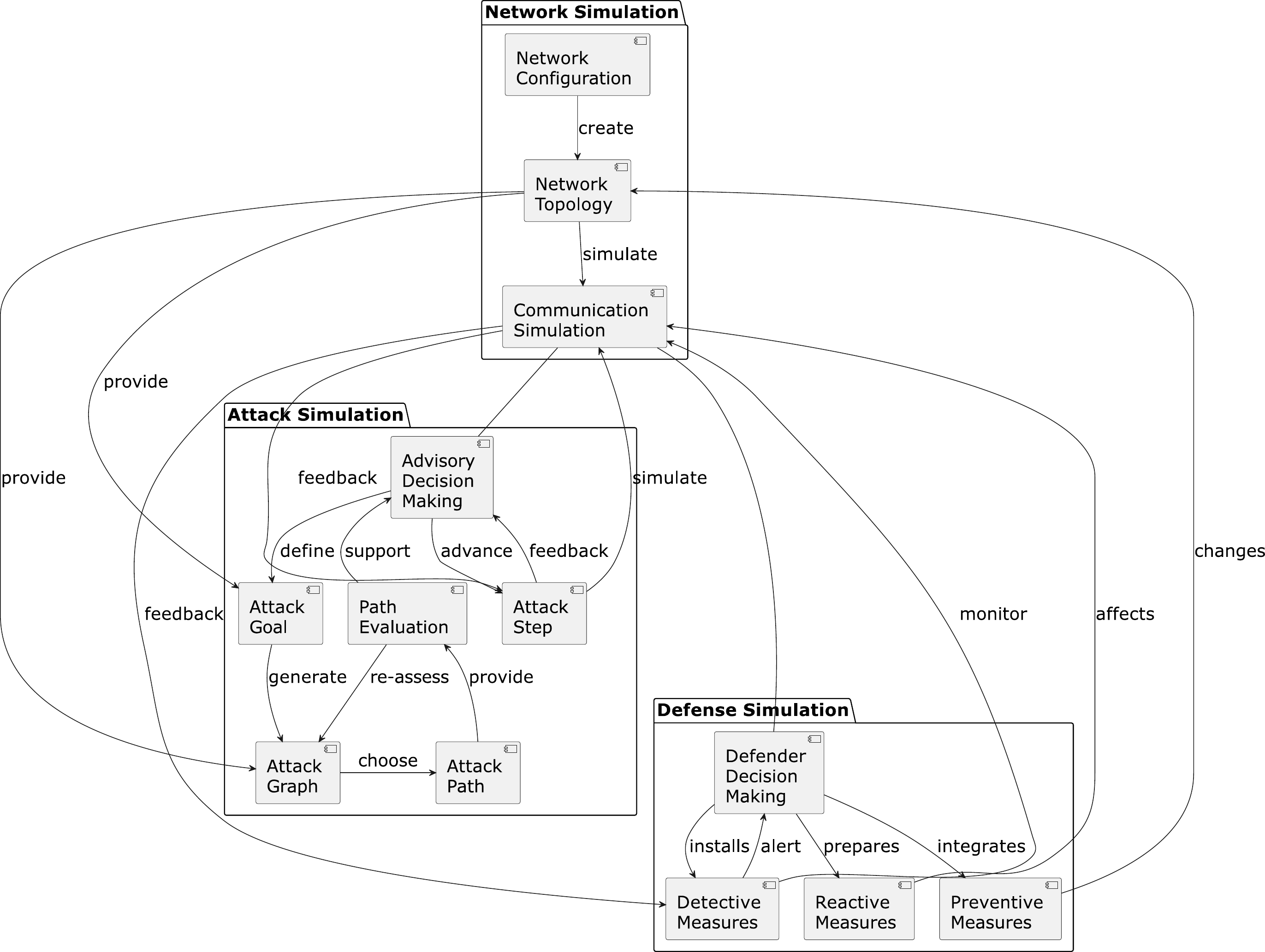}}
\caption{A comprehensive representation of our approach, highlighting a game-theoretic model that simulates the dynamic interactions in cyber warfare, emphasizing ongoing learning and strategic adjustments.}
\label{fig:framework_overview}
\end{figure*}
Our approach is rooted in game theory to examine the dynamics involved in cyber warfare, as illustrated in Figure~\ref{fig:framework_overview}.
We simulate a confrontation scenario, often termed red team vs.\ blue team, to assess how these interactions impact the quality of data.
In this simulation, the attacker's goal is to disrupt grid operations, and the defender's goal is to thwart these efforts.
The attacker methodically assesses the infrastructure to identify targets with significant potential for outage costs. Conversely, the defender implements strategies to mitigate these risks.
Our model features dynamic attack graph generation, which evolves with each iteration of the game, taking into account the defensive measures. 
The attack’s likelihood of success is determined using historical data, with the attacker selecting the most efficient route via Dijkstra's Algorithm.
Learning capabilities are assigned to both the attacker and defender, influencing their tactics as the simulation progresses. This iterative learning process involves transferring insights from one round to the next.

Essential network nodes in our framework are secured with \ac{IDS}, which triggers alerts to possible threats within a Python environment. This setup is supported by executing \ac{MulVAL}~\cite{b10} within Docker containers.
\ac{ML} models, specifically developed using Python libraries such as sklearn~\cite{hao2019machine}, are refined with Grid Search for optimal hyperparameter selection, aiming to reduce overfitting.

\subsection{Multi-Stage Attack Modeling}
To accurately simulate attacks on complex infrastructures such as power grids, it is essential to consider multi-stage attacks rather than merely single-point breaches.
Tools for generating attack graphs are adept at mapping out complex scenarios where multiple vulnerabilities are exploited simultaneously, resulting in sophisticated, multi-tiered cyber attacks.
These tools consider a range of factors, including the operating environment, the severity of vulnerabilities, and their consequential impacts.
A variety of these tools were evaluated. We observed that open-source options, despite their complexity and less intuitive visual presentations, offer more comprehensive insights into potential attack pathways. Thus, for a balance of detail and scalability, we selected tools such as \ac{MulVAL} for their robustness and adaptability.

\ac{MulVAL}, a tool focused on creating logical attack graphs, is notable for its use of Datalog as an input syntax. This choice facilitates the integration of established vulnerability databases such as the \ac{NVD}, enabling specific vulnerability identification through their unique IDs.
Additionally, it allows for the integration of detailed host and network configurations, obtainable through an \ac{OVAL} scanner and firewall management systems, enriching the contextual understanding of potential attacks.

\subsection{Dynamics of Cyber Attack and Cyber Defense}\label{subsec:attacker_defender}
\begin{figure}
\centerline{\includegraphics[width=\columnwidth]{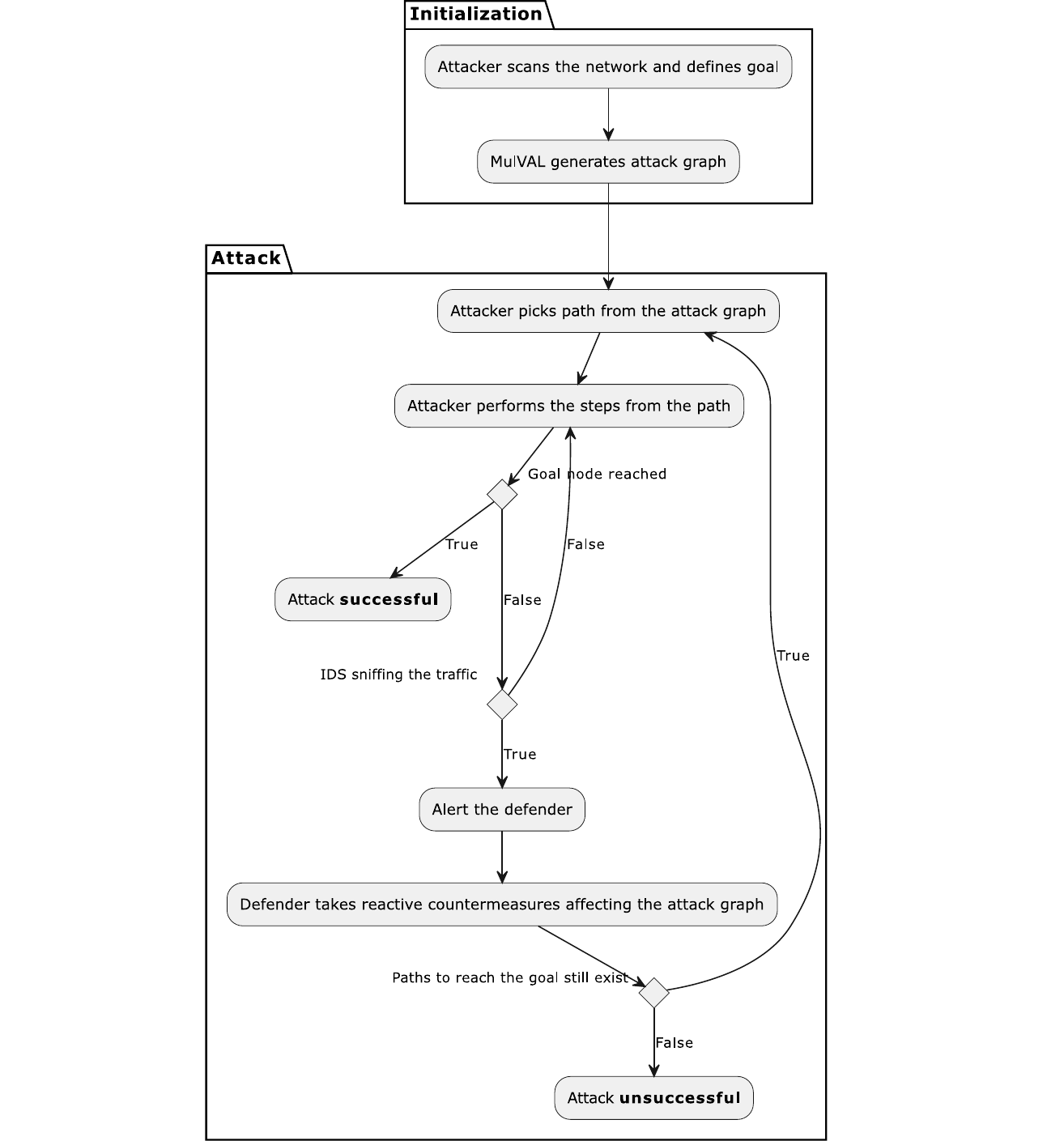}}
\caption{Depiction of the interactive dynamics between attacker and defender, illustrating how the attacker probes the network and the defensive measures put in place, including the usage of signature-based \ac{IDS} sensors within the simulation for detection.}
\label{fig:attack_defense}
\end{figure}

In our analysis, we apply a game-theoretical model to characterize the interaction between attackers and defenders, as suggested in~\cite{b13}.
This interaction is depicted in Figure~\ref{fig:attack_defense}.
We employ the frameworks of MITRE ATT\&CK~\cite{strom2018mitre} and D3FEND~\cite{kaloroumakis2021toward} to model the behavior of both parties.
The attacker begins by scanning the network to pinpoint the most vulnerable node, often those associated with high potential outage costs, such as a \ac{SCADA} Server, which is particularly prone to costly disruptions due to its high replacement cost and critical role in grid stability.

The defender's strategy encompasses both proactive and reactive steps to mitigate these risks, following the guidelines of~\cite{kaloroumakis2021toward}.
Methods include the deployment of \ac{IDS} for monitoring and access control mechanisms for prevention.
In the simulation context, \ac{IDS} sensors, specifically signature-based ones, are deployed to assess the system's defense capabilities by identifying anomalous activities based on established rules.
These sensors serve as a measure of the defense system's effectiveness in the simulation and are not factored into the final evaluation stage, where the focus shifts to \ac{ML}-based anomaly detection metrics.
The simulation experiments with varying numbers of sensors to gauge their impact on data generation.

We also utilize \ac{MulVAL} to generate an attack graph that mirrors the attacker's course of action.
The attacker maneuvers through the network, compromising elements until either achieving their objective or being detected by an \ac{IDS}, which triggers the defender's responsive actions.
The encounter continues until either the attacker attains their goal or the paths available become impracticable, signifying a thwarted attack.
Experience from each encounter informs both the attacker and defender, leading to enhanced strategies and more sophisticated attacks in subsequent simulations.

\subsection{Offensive Strategy in Cyber Attacks}
\label{subsec: Attacker}
The primary objective of the attacker is to disrupt grid operations, while strategically avoiding detection by \ac{IDS}.
The extent of disruption is assessed based on the outage costs resulting from the attacker's maneuvers.
The attacker progressively enhances their expertise, elevating their proficiency level with each successive attack. This progression also includes updating their understanding of the probability of successfully breaching a node.
This proficiency, or skill rate, is a key determinant in the success of an attack and is influenced by both previous successes and failures, following a probabilistic model.
These elements guide the attacker in selecting their course of action, employing Dijkstra's algorithm to identify the most effective path, as referenced in~\cite{dijkstra1959note}.
The decision-making process involves calculating edge $(i,j)$ weights in the attack graph, as shown in Equation~\ref{eq:edge_weight}:

\begin{equation}
W_{i, j} = \frac{t_{j}^{attacker}}{C_{j}^{attacker} \cdot P_{j}^{attacker}}
\label{eq:edge_weight}
\end{equation}

In this equation, the ``attacker'' superscript signifies the evaluation perspective. The term $t$ signifies the time required for a specific action, as defined by the \ac{TTC} metric detailed in Section~\ref{subsec: Risk evaluation}.
$P_{j}^{attacker}$ quantifies the attacker's estimated success probability in compromising a node.
At the outset, the attacker presumes a default success rate of 1. However, with each attempt, whether successful (1) or not (0), this rate is recalibrated based on the aggregate success history.
A successful breach is determined by comparing a randomly generated number with the calculated success rate; if unsuccessful, the attacker either retries, thereby increasing resource consumption and prompting the defender to bolster their defenses, or re-evaluates their chosen path.
The term $C_{j}^{attacker}$ represents the outage costs from the attacker's viewpoint.

\subsection{Defensive Strategies in Cyber Security}
\label{subsec: Defender}
The primary objective of the defender is to reduce the likelihood of disruptions to grid operations caused by cyber attacks.
Their strategy includes both proactive and reactive measures: proactive measures necessitate financial investment and are implemented over multiple simulation rounds, while reactive measures are more immediate, requiring no additional resources within the same simulation round.
The defender's budget, encompassing both initial and accumulated funds, is allocated towards implementing these cyber security measures.
By analyzing the attacker's most used paths from past attacks, the defender continuously refines their understanding and approach to risk management.
This analysis directly influences the risk assessment and the design of proactive defenses.
Risk evaluation is quantified in Equation~\ref{eq:risk_learn}, where a learning rate, $Q_{i}$, is assigned to each node $i$ in the network. This rate starts at 1 and increases with each detected attack:

\begin{equation}
Risk = \sum_{i = 1}^N P_i \cdot C_i \cdot Q_{i}
\label{eq:risk_learn}
\end{equation}

For effective deployment, \ac{IDS} sensors are strategically placed at nodes with higher potential outage costs.
To identify critical network nodes, centrality algorithms, particularly the current flow betweenness centrality method, are employed. This method considers the network as an electrical circuit, differing from traditional betweenness centrality that assumes linear information flow.

\begin{equation}
c_{CB}(v) = \frac{1}{(n - 1) \cdot (n-2)} \sum_{s, t \in V} \tau_{st} (v)
\label{eq:current_flow_betw_centr}
\end{equation}

\begin{equation}
\tau_{st} (v) = \frac{1}{2}( - |b_{st}(v)| + \sum_{e \ni V} |x(\vec{e})|)
\label{eq:throughput}
\end{equation}

Equation~\ref{eq:current_flow_betw_centr} computes the normalized current passing through a node $v$, with $\tau_{st}(v)$ representing the throughput and a normalization factor. Equation~\ref{eq:throughput} calculates the current using $b_{st}(v)$, balancing it across all $b_{st}(v)$ in the network while also factoring in edge resistances as part of the centrality calculation.

\begin{equation}
r(\vec{e_{i, j}}) = \frac{1}{\max(c_i ^ {outage}, c_j ^ {outage})}
\label{eq:resistance}
\end{equation}

The resistance of each edge, defined in Equation~\ref{eq:resistance}, plays a crucial role in determining centrality, with lower resistance along edges between nodes with higher outage costs. As the simulation progresses, the defender adapts their sensor placement, learning from the attacker's previous actions and modifying their defense strategy accordingly.

\subsection{Guiding Decision-Making}
\label{subsec: Risk evaluation}
In the game-theoretical framework of attacker-defender interactions, a benchmark value is essential for assessing the effectiveness of their strategies.
This evaluation is conducted using risk assessment models tailored for cyber attacks on \ac{SCADA} systems, which encompass aspects such as vulnerability, threat dynamics, defensive actions, and their overall impact as discussed in~\cite{b18}.
From the defender's point of view, this risk value is crucial to minimize, aiming to avert or lessen potential harm. Conversely, the attacker seeks to maximize this value to achieve their objectives.
The $\beta$-\ac{TTC} Metric, which plays a pivotal role in practical cyber security risk estimation, takes into account both system vulnerabilities and the skill level of the attacker, providing an estimate of the time required for system compromise~\cite{b19}.
This metric, along with the success probability for each component of the attack $P_j$, is calculated following the methodology in~\cite{b19}, which leverages vulnerability information from public databases such as the \ac{CVE}, aligning well with \ac{MulVAL} criteria.

\begin{equation}
TTC(s, W) = t_1 \cdot P_1 + t_2 \cdot (1 - P_1) \cdot (1 - u)
\label{eq:bttc}
\end{equation}

The impact factor, represented by each component's outage costs $C_i$, is evaluated based on its significance in grid functionality.
These outage costs are determined using methodologies such as the Purdue model~\cite{b6} and an industry-specific model accounting for the cost of unplanned outages based on peak power consumption~\cite{b22}.
In our model, we consider a severe scenario of a 12-hour grid outage.
Risk evaluation is then articulated using Equation~\ref{eq:risk}:

\begin{equation}
Risk = \sum_{i = 1}^N P_i \cdot C_i
\label{eq:risk}
\end{equation}

Through this interactive model, a sequence of events in a cyber incident is simulated, reflecting the complex dynamics between attackers and defenders. This simulation forms the basis for generating synthetic data, capturing varied scenarios of cyber incidents.

%% file: chapter4d.tex
\section{Laboratory Tests} \label{sec:labres}
Conducting laboratory tests is vital for validating the effectiveness and realism of the simulation environment. This section describes the experimental setup and methodologies used in the laboratory to evaluate the implemented solutions in real-world conditions. While we conducted our physical tests in our smart grid laboratory in Aachen, our research setup for the simulation utilized a high-performance PC equipped with a multicore CPU, a minimum of 16GB of RAM, a dedicated GPU with at least 8GB of VRAM, and SSD storage.

\subsection{Experimental Setup} \label{subsec:labres_setup}
\begin{figure*}
\centerline{\includegraphics[width=\linewidth]{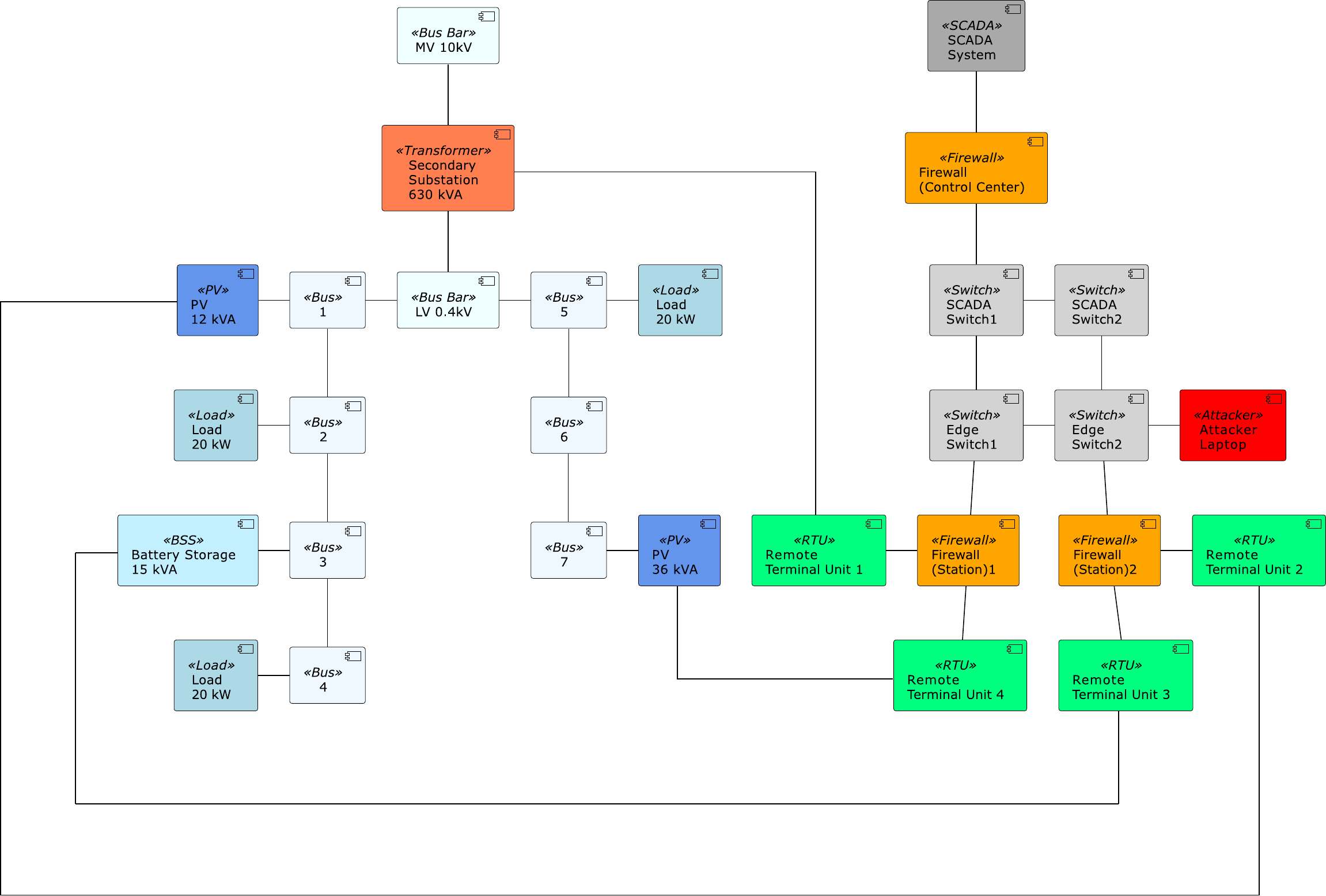}}
\caption{Overview of the laboratory setup illustrating an MV/LV grid with battery storage systems, PV inverters, and load banks. These components are remotely controllable and monitored through a local \ac{SCADA} system.}
\label{fig:labres_setup}
\end{figure*}

The evaluation of the implemented solutions follows a step-by-step approach, conducted in both a laboratory and a simulation environment. The experimental setup, as illustrated in Figure~\ref{fig:labres_setup}, replicates a low-voltage grid with various consumers and producers. The experiment is based on prior works~\cite{sen2021investigating}.

The setup includes three thermal loads, each with a maximum power consumption of 20 kW, representing a residential street.
Additionally, three facilities for feeding electrical energy are used: A 12 kW and a 36 kW solar inverter and three 5 kVA battery inverters (total 15 kVA).
The grid is connected to the medium-voltage grid through a local secondary substation to ensure supply security.
All feeding plants are accessible remotely via control devices and a communication network from \ac{SCADA}.
The network architecture employs multiple switches for redundancy, with sophisticated firewalls and switches capable of handling maximum data flow. The \ac{L3} participants in the network, however, are more rudimentary. A Raspberry Pi 4 is used for the control center, and the \acp{RTU} are connected via 100BASE-T Ethernet and tasked with relaying control commands.
A commercially available laptop is used to simulate an attacker's position on an Edge Switch. Network packets are captured and analyzed using another laptop equipped with tshark (Version 3).

To examine various attack scenarios on a power grid, the results from real experiments are compared with those from the simulation. Parameterization of the model is carried out by measuring data rate and latency. Two key measurements are conducted:

\begin{itemize}
    \item Net data rate and latency for Gigabit-Ethernet are measured between two laptops connected by a switch. iPerf3 (Version 3.9) and ICMP ping tests are used. Results show a data rate of 941 Mbit/s and an average \ac{RTT} of $343 \mu s$.
    \item Another measurement is performed with an ICMP ping from a laptop to the control device of a 36 kW solar inverter. The average \ac{RTT} is approximately $540 \mu s$.
\end{itemize}

The results of these measurements are used to adjust the parameters of the network model to make the simulation as realistic as possible. The modeling considers the maximum achievable transmission speed and delays in data transmission, derived from the measurement results. This ensures an accurate representation of the real network conditions in the simulation.

For effective management, the local \ac{SCADA} system controls the energy supply, aiming to maximize local sources and minimize dependence on the medium-voltage network, i.e. self-consumption optimization. The attacker, on the other hand, seeks to disrupt this communication to compromise supply security or increase operational costs.

\subsection{Results}\label{subsec:labres_res}
\begin{figure*}
\centerline{\includegraphics[width=\linewidth]{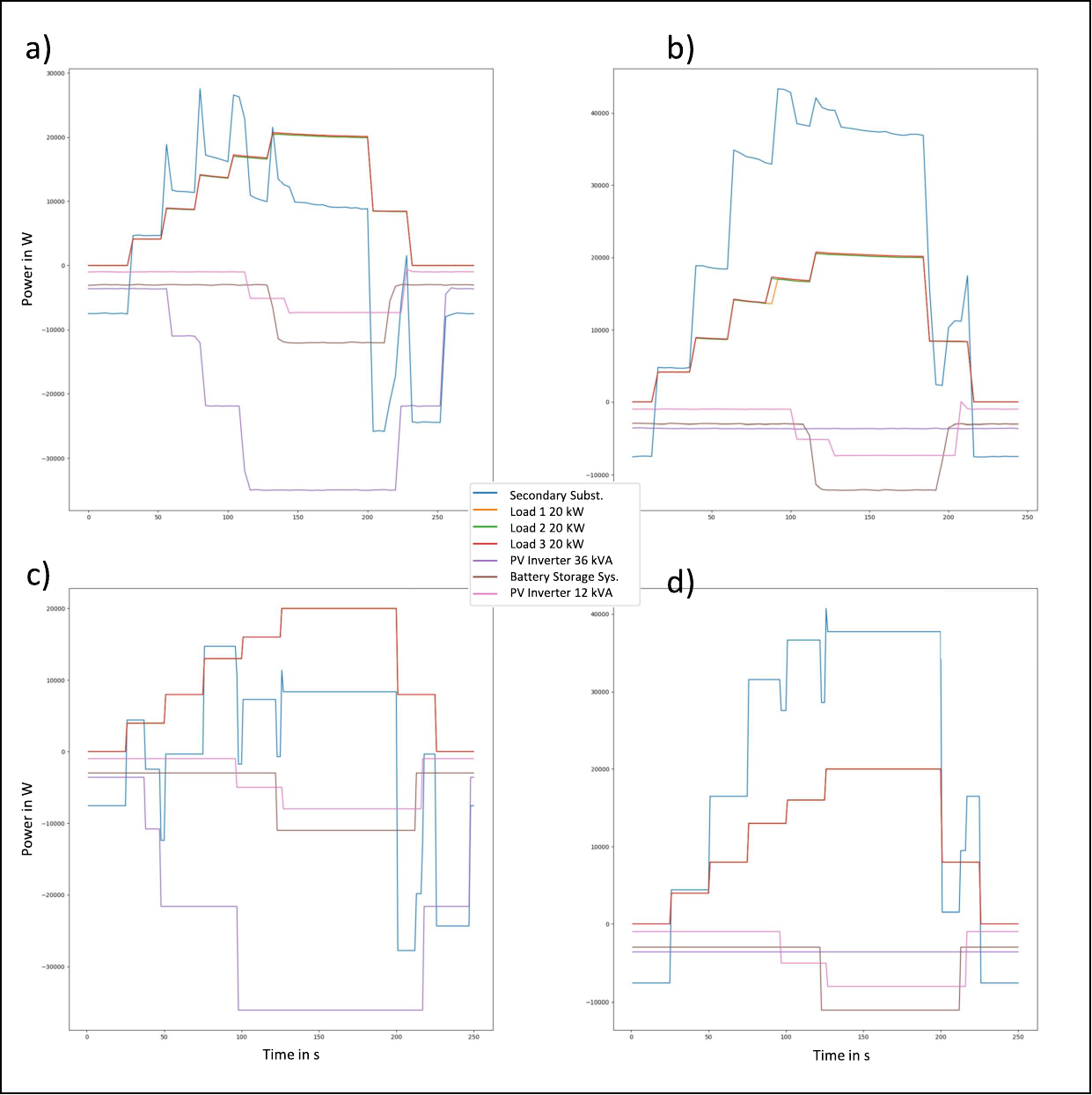}}
\caption{Overview of the results from the experiments conducted in both the laboratory and the simulation environment. Plot a) represents the normal scenario in the laboratory, and plot c) depicts the corresponding scenario in the simulation environment. Plot b) illustrates the ARP spoof-based attack scenario in the laboratory, while plot d) shows the same in the simulation environment.}
\label{fig:labres_results}
\end{figure*}

To ascertain the fidelity of the simulation results and to validate the simulation environment, four distinct scenarios were examined both in the laboratory and within the simulation environment (cf. Figure~\ref{fig:labres_results}). These scenarios were not designed to provide intrinsic value in energy technology or information security but rather to demonstrate the feasibility of conducting serious scientific investigations.

The initial scenario explored the system's behavior under standard regulation conditions without an attacker, focusing on the response of the control center to increasing power demand. This was followed by three distinct cyber attack scenarios, including an ARP spoofing attack and two ICMP flooding attacks. 

The laboratory results are presented in a series of diagrams in Figure~\ref{fig:labres_results} showing power consumption and injection for the seven electrical installations over the duration of the experiments, measured in seconds. This included three thermal load,  two PV Inverters, a battery inverter, and an adjustable local secondary substation. The power balance resulting from the first six installations equates to the compensation that secondary substation will provide via the medium voltage network to ensure grid stability.

In the absence of an attacker, the control center's response aimed to cover the increasing load effectively. However, the ARP spoofing and ICMP flooding attacks revealed significant discrepancies in the secondary substation's power trajectory, with consumption sometimes exceeding 40 kW, contrasting with peaks of under 30 kW during normal operations. 

The ARP spoofing scenario, executed via a laptop connected to an Edge-Switch, employed Ethercap under Linux targeting the \ac{SCADA} system, resulted in the control center losing communication with the remote terminal unit of the PV Inverter 36 kVA. 

For the ICMP flooding attack, hping3 was utilized, inundating the remote terminal unit with approximately 120,000 queries per second. This deluge led to filled network buffers and packet loss, thereby preventing the \ac{SCADA} system from establishing TCP connections and disrupting communication with the remote terminal units.

Simulated scenarios exhibited promising concordance with laboratory outcomes. Although slight variations were evident due to the physical delays inherent in real-world devices, the overarching behavior was consistent with empirical data. The simulation confirmed that the communication modeling within the framework could accurately reflect the control center's interaction with the power installations.

These experiments underscore the simulation environment's ability to mirror real-world operations with a high degree of realism. The replication of cyber attack impacts in the simulation further validated the framework's efficacy in modeling such disruptions and their potential consequences on grid stability and control center operations.

\subsection{Discussion}\label{subsec:labres_dis}
Our simulation environment offers a significant leap toward integrating energy and communication systems into a singular application, a step forward from co-simulations that often impose complexity. 
It is designed to lower the barrier to entry, enabling users without deep knowledge in communication technology to engage in simulations effectively.

The Ethernet and WLAN reference implementations serve as vital examples, providing a foundation for users to build upon. They simplify the process for new users and demonstrate potential expansion paths for various investigations. This user-friendliness extends the practical reach of the simulation framework, making it a valuable tool for researchers.

However, the simulation's accuracy depends on the fidelity of the models used. Theoretical models may not always align with real-world behavior, as seen in the laboratory's empirical tests. Thus, to achieve the most accurate simulations, it's advisable to calibrate models against physical measurements rather than relying solely on literature values.

Despite these limitations, the simulation framework's structured approach, with defined interfaces and tasks, presents an advantageous alternative to network simulators, offering clear guidance for implementing communication technologies. Future work could focus on expanding the framework to encompass a broader range of these technologies, further enabling the exploration of complex power grid scenarios within the simulation environment.

The simulation environment’s integration of energy and communication systems emphasizes the development of an accessible and comprehensive simulation framework for power grid cyber security. By lowering the barrier to entry and providing reference implementations such as Ethernet and WLAN, the framework democratizes the ability to simulate complex cyber physical interactions, making it a powerful tool not only for experts but also for researchers and practitioners less familiar with communication technology. This ease of use ensures that a wider audience can engage in the critical task of simulating and addressing vulnerabilities in smart grid systems. Moreover, the structured nature of the simulation, with well-defined interfaces and tasks, enhances its scalability and flexibility, making it suitable for a range of investigations into grid resilience. While real-world accuracy remains a challenge due to the reliance on theoretical models, the framework’s ability to serve as a bridge between co-simulation complexity and real-world application strengthens its role as a foundational tool for future research and development in power grid cyber security. This contribution to making smart grid simulations more accessible and practical underpins the paper’s broader objective of advancing security solutions for critical infrastructure.

%% file: chapter4b.tex
\section{Attack Simulation Test} \label{sec:resatt}
The testing of the simulation phases plays a key role in ensuring the accuracy of the data samples used for \ac{ML}-based \ac{IDS} training. To generate reliable and representative data, it is crucial to first validate the plausibility of the simulated attack scenarios. This involves thoroughly testing the simulation environment to ensure that both the physical behavior of the power grid and the communication processes under normal and attack-induced conditions are accurately replicated. Once the simulation's credibility is established, it is further evaluated by comparing its results with those obtained from a real cyber physical laboratory system, where similar attack scenarios are replicated. This comparison verifies the precision of the simulation in reflecting real-world behaviors, ensuring that the synthetic data produced for \ac{IDS} training is both realistic and comprehensive. Ultimately, this process guarantees that the \ac{IDS} models can effectively detect and respond to cyber attacks, making the data generation a critical component in the overall security framework.

In this section, we present case studies that offer comprehensive insights into the practical applications of the attack simulation framework, illustrating its adaptability and effectiveness in various scenarios representative of real-world grid vulnerabilities and cyber attack patterns.

\subsection{Procedure} \label{subsec:resatt_proc}
The investigation procedure for attack simulation verification focuses on analyzing the impact of input parameters on the modeled attacker, consisting of the Attack Propagator (simulating the \ac{IT} compromise of devices) and the Coordinator (executing energy-related actions regarding the attack). Parameters influencing the propagation behavior of the attacker and, indirectly, the actions of the Coordinator are examined.

Key varying parameters include the firewall configuration and attacker metadata. These directly affect the possibilities of attack propagation and the coordination of attacks on the energy technology.

\begin{itemize}
    \item \textbf{Firewall Configuration:} This parameter determines the virtual zonal segmentation and hierarchical connection routes, thus influencing the distribution of vulnerabilities in the grid. The assignment is zone-specific, based on the assumption that remotely located devices are easier to compromise.
    \item \textbf{Attacker Metadata:} This includes information such as privileged accesses, range, and grid file system mounts, which are defined via vulnerability vectors and assigned to devices.
\end{itemize}

The attack simulation is carried out by selecting a measurement value to be manipulated (e.g., active power at a substation), followed by the calculation and application of an optimal attack vector. This manipulation influences the results of state estimation in the grid and the control commands based on it.

\begin{enumerate}
    \item \textbf{Definition of the Network Scenario:} Selection of grid topology and state variables, based on real or hypothetical grid configurations.
    \item \textbf{Application of the \ac{FDI} Attack:} Manipulation of the measurement values and analysis of the effects on the estimated grid state and operational control.
    \item \textbf{Comparison of Grid States:} Investigation of the differences between the real grid state with and without \ac{FDI} attack, and the state simulated by the attack.
\end{enumerate}

Through this investigation, the impact of measurement value changes on operational control logic and the risk of critical grid states can be assessed.

\subsection{Case Studies} \label{subsec:resatt_res}
This section presents the results of the case studies, focusing on verifying the attack simulation aspect of the simulation environment.

\begin{figure*}[ht]
\centerline{\includegraphics[width=\columnwidth]{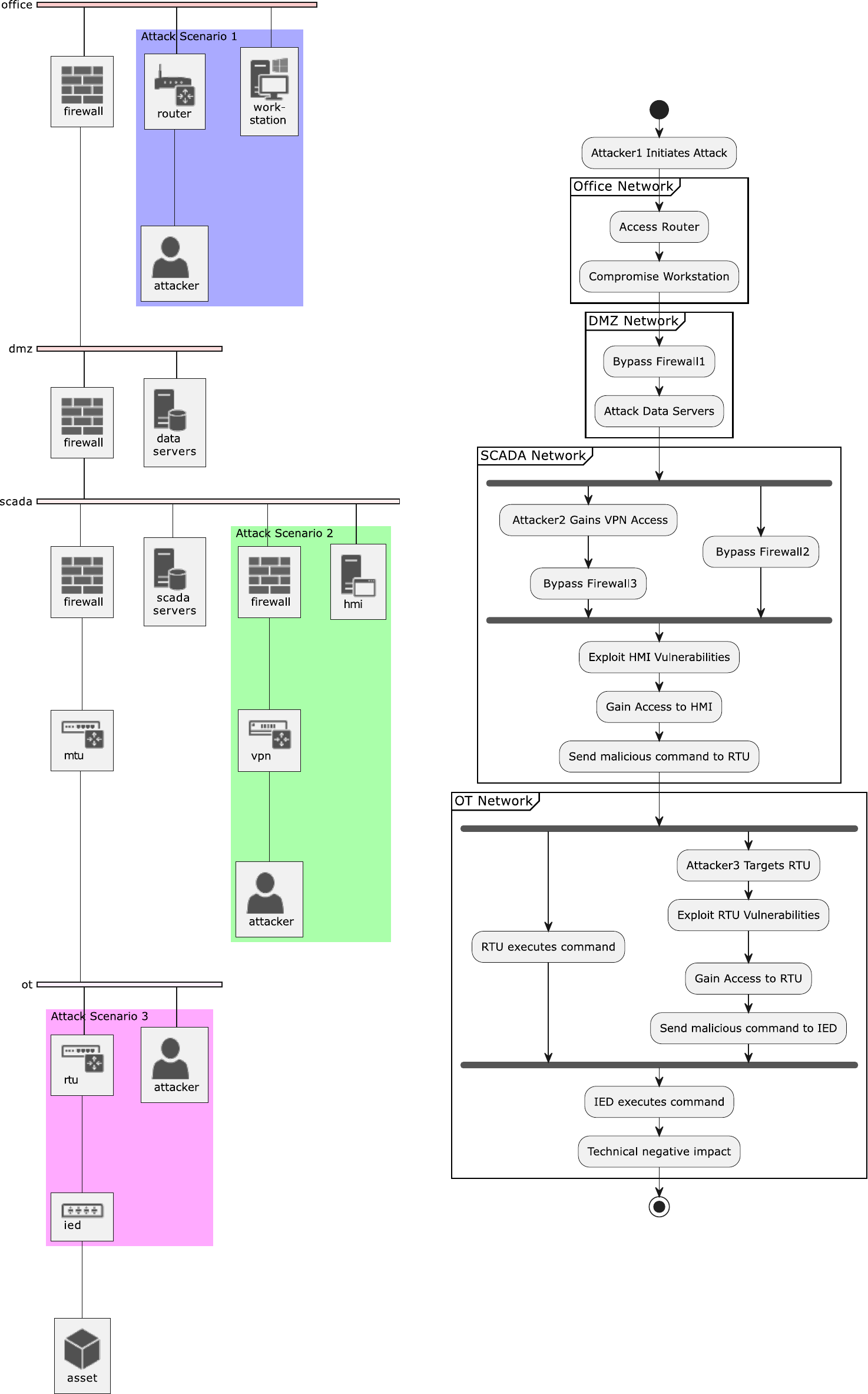}}
\caption{Illustration of the simulated attack propagation within the simulation environment with respect to different scenario constellations.}
\label{fig:att_comb}
\end{figure*}
\paragraph{Attack Scenario 1}
In this scenario (cf. Figure~\ref{fig:att_comb}, attack scenario 1), Attacker1 initiates the attack at the field level, directly compromising a workstation within the Office Network. Progressing vertically, the attacker bypasses Firewall1 to attack data servers located in the \ac{DMZ} Network. Subsequently, Attacker1 exploits vulnerabilities in the \ac{SCADA} Network's \ac{HMI} and sends malicious commands to an \ac{RTU} in the \ac{OT} Network. This vertical propagation emphasizes the attacker's ability to move up through the network layers to reach operationally critical systems.

In this scenario, the lack of virtual zonal segmentation means that the attacker faces fewer barriers in moving from the field level to the operation level. However, the strict hierarchical connections imposed by the firewall configuration mean that the attacker must follow a defined path, encountering each node's specific defenses. The attacker metadata, such as gained privileges and access range, dictate the effectiveness of each action, with higher privileges enabling deeper penetration into the network.

\paragraph{Attack Scenario 2}
Diverging from the first scenario, Attacker2 exhibits both vertical and lateral propagation abilities (cf. Figure~\ref{fig:att_comb}, attack scenario 2). After successfully gaining VPN access, Attacker2 focuses on Firewall3. If they successfully bypass it and gain access to the \ac{SCADA} servers, they proceed to exploit these servers. Concurrently, the attacker also attempts to bypass Firewall2. Successful bypassing leads them to the \ac{HMI}. The attacker then exploits vulnerabilities within the \ac{HMI} to gain control. The final objective in this stream is to send malicious commands to the \ac{RTU}. 

The absence of access control mechanisms opens up the network, allowing the attacker to propagate laterally within the \ac{SCADA} zone. The ability to navigate through and exploit vulnerabilities in such an environment reflects the sophisticated and targeted nature of the attacker. In this scenario, the attacker’s lateral movement capability is a direct result of the more relaxed firewall configuration without enforced access control rules, demonstrating how network defenses significantly influence the attacker's reach.

\paragraph{Attack Scenario 3}
In the third attack scenario (cf. Figure~\ref{fig:att_comb}, attack scenario 3), Attacker3 showcases a combination of both vertical and lateral propagation abilities. Starting from a station-level node, the attacker exploits the lack of hierarchical connections (horizontal communication allowed), facilitating movement within the \ac{WAN} zone. This scenario highlights the attacker's capability for lateral movements across different network segments. The absence of stringent hierarchical connections in a flat network topology, especially without robust firewall rules, opens up possibilities for widespread compromise. The attacker’s ability to move laterally is significantly influenced by the more relaxed firewall configurations, underscoring the critical role of network defenses in controlling an attacker's reach.

Concurrently, this scenario also illustrates the effect of enforced zonal segmentation by firewalls on attacker movements. Specifically, Attacker3, beginning their campaign at an \ac{RTU} within the \ac{OT} network, is confined to the station level. This confinement is a strategic outcome of the firewall-imposed virtual zonal segmentation. The attacker's primary objective is to maximize the number of compromised devices within this zone, eventually triggering a critical state in the network. However, the robust network perimeter defenses significantly limit the attacker's movements. This scenario, therefore, demonstrates the effectiveness of strategic firewall deployment in containing an attack, restricting its movement to certain network segments, and consequently mitigating potential damage.

\paragraph{Attack Scenario 4}
Building on the constraints of Scenario 3, additional hierarchical connection structures further restrict the attacker's movements such as no horizontal communication is allowed (cf. Figure~\ref{fig:att_comb}, attack scenario 3). The attacker, starting at an \ac{RTU}, is unable to progress the attack, representing a scenario where the network operates under normal conditions without any attack progression. This scenario underscores the efficacy of a well-structured network in mitigating the risk of an attack.

Adding a hierarchical connection structure parameter, the simulation shows an environment where the attacker's movement is severely restricted. This scenario validates the effectiveness of a defense-in-depth strategy. Even if the attacker compromises a station-level device, the inability to move either up or down the network hierarchy prevents any further attack progression. The attack is stifled due to both vertical and horizontal movement restrictions.

\subsection{FDI Attack} \label{subsec:resatt_fdi}
This study investigates the implications of \ac{FDI} attack tactics on Simbench medium-voltage networks. It explores two principal scenarios: the concealment of an actual overload problem and the fabrication of a non-existent overload problem to elicit unnecessary grid regulatory actions.

The simulation modeled a typical medium-voltage grid topology with the following critical operating intervals:

\begin{itemize}
    \item Voltage levels between 0.965 pu and 1.055 pu.
    \item Maximum line loading capped at 100\%.
\end{itemize}

The grid is visualized as an open ring grid with congestion issues introduced by high consumer demand on specific branches.

The study carried out two experiments:

\begin{enumerate}
    \item \textbf{Overload Problem Obscuration:} Manipulation of active power measurements to hide an overload issue at substation SS 73, artificially increasing it by 3 MW (cf. Table~\ref{tab:attack_vector_3MW}).
    \item \textbf{Feigned Overload Problem:} Creation of a false overload situation by increasing the power flow reading by 4 MW on the line from SS 62 to SS 63 leading to a maximum loading of 109.40 \% at Line 1 - 2 (cf. Table~\ref{tab:attack_vector_4MW}).
\end{enumerate}

\begin{table}[htbp]
\centering
\caption{Attack vector for increasing 3 MW at Busbar SS 73}
\label{tab:attack_vector_3MW}
\begin{tabular}{p{0.75cm}p{1.5cm}p{1.5cm}p{1.5cm}p{1.5cm}}
\toprule
Meas. \( k \) & Element Type & Attack Vector \( a_k \) & Correct Value [MW], \( z_k \) & Falsified Value [MW], \( \hat{z}_k \) \\
\midrule
67 & SS 72 & -1.6860 & 0.272 & -1.4140 \\
68 & SS 73 & 3.00 & 0.272 & 3.272 \\
69 & SS 74 & -1.3139 & 0.2365 & -1.0774 \\
212 & Line 72 – 73 & 1.6860 & 1.1296 & 2.8156 \\
213 & Line 73 - 74 & -1.3145 & 0.8574 & -0.4571 \\
\bottomrule
\end{tabular}
\end{table}

\begin{table}[htbp]
\centering
\caption{Attack vector for increasing 4 MW on Line 62 - 63}
\label{tab:attack_vector_4MW}
\begin{tabular}{p{0.75cm}p{1.5cm}p{1.5cm}p{1.5cm}p{1.5cm}}
\toprule
Meas. \( k \) & Element Type & Attack Vector \( a_k \) & Correct Value, \( z_k \) & Falsified Value, \( \hat{z}_k \) \\
\midrule
57 & SS 62 & -4.00 & 0.237 MW & -3.763 \\
58 & SS 63 & 4 & 0.384 MW & 4.384 \\
202 & Line 62 - 63 & 4.00 & 4.794 MW & 8.794 \\
\bottomrule
\end{tabular}
\end{table}

The grid's response to the \ac{FDI} attack was twofold:

\begin{enumerate}
    \item When no \ac{FDI} attack occurred, the grid operator effectively returned the grid to normal operation through topology switches.
    \item Under \ac{FDI} attack, the critical grid state was either successfully obscured or falsely represented, influencing the grid operator's response to a non-existent condition.
\end{enumerate}

The attacker's success hinged on the capacity to analyze the current grid state and identify exploitable scenarios. The assumption that the attacker had compromised relevant RTUs may not hold in a secure grid environment.

\subsection{Discussion} \label{subsec:resatt_dis}
Our simulation results highlight the critical role of parameterization in dictating the trajectory and impact of cyber attacks within network environments. The interaction between firewall configuration and attacker metadata shapes the landscape of vulnerabilities and potential attack pathways.

Firewalls serve as checkpoints in the network, determining data flow and accessibility. Their configuration plays a pivotal role in shaping the attacker’s access points and routes. Virtual zonal segmentation and hierarchical connections introduce complexity and highlight potential vulnerabilities of field devices.

This parameter sets the attacker's skill level, reach, and method of exploitation. Vulnerability vectors assigned to devices measure each node's resilience against the attacker’s capabilities, simulating a range of attack strategies.

The process involves defining the grid scenario, executing \ac{FDI} attacks, and comparing simulated and actual grid states. This approach offers insights into the effectiveness of defense mechanisms and the realism of the simulation.

While the simulation provides valuable insights, it is crucial to acknowledge its limitations. In simulations, outcomes can be biased by parameter selection and may not capture the complexity of real-world networks, including dynamic changes and unpredictable user behavior. Additionally, the predictability of attacker behavior and the rapid evolution of cyber threats pose significant challenges, often resulting in oversimplifications that fail to mirror actual conditions.

Regarding the \ac{FDI} attack scenarios, the simulations demonstrated grid state estimation's susceptibility to \ac{FDI} attacks. The attacker could mask or simulate grid issues, though focused attacks on specific branches rather than the entire grid sufficed.

The adherence of the grid to the (n-1) criterion illustrated some resilience to manipulation. However, the successful obscuration of real problems highlights a significant risk to grid operations, emphasizing the need for robust detection mechanisms.

The parameterization within cyber attack simulations emphasizes the importance of a realistic and adaptable simulation environment for understanding and mitigating cyber threats in smart grids. The interplay between firewall configurations and attacker capabilities highlights the dynamic complexity of the simulated environment, demonstrating how defense mechanisms can influence the trajectory of attacks. By adjusting parameters such as vulnerability vectors, the simulation offers a range of attack scenarios, providing valuable insights into the network's defense strategies and how they can be enhanced to meet real-world challenges. This flexibility in simulating varying attack paths and defense mechanisms underlines the simulation environment's role in generating diverse, high-quality datasets for IDS training, ensuring that these models are exposed to a broad spectrum of cyber threats. The recognition of limitations in parameter selection and real-world unpredictability further emphasizes the need for continued refinement, aligning with the paper's overall objective to develop a robust and dynamic framework capable of addressing the evolving landscape of smart grid cyber security. The insights drawn from these attack and defense interactions validate the simulation’s role in improving detection systems and providing a more comprehensive understanding of grid vulnerabilities.

%% file: chapter4a.tex
\section{Synthesizing Data for IDS} \label{sec:result}
The detailed simulation of our proposed approach also enables the capability to generate synthetic data sets for training \ac{ML}-based \ac{IDS}. This data generation plays a foundational role in enabling the simulation of complex, multi-layered cyber attacks, such as \ac{FDI} and denial-of-service attacks, across 21 sub-networks. The environment produces high-quality, diverse data samples that incorporate real-time attacker-defender dynamics, allowing for the development of \ac{IDS} models that can effectively detect and respond to evolving threats. An essential feature of this environment is its ability to generate synthetic data that represents critical infrastructure under attack, which is difficult to obtain in real-world scenarios. This not only enables the investigation of critical security incidents but also supports the holistic security approach of prevention, detection, and reaction by providing the necessary datasets to train and evaluate \ac{IDS} approaches. This capability ensures that \ac{IDS} models are well-equipped to handle advanced cyber threats, contributing to the overall resilience of smart grid systems.

In the following, we summarize our methodology for \ac{IDS} training using \ac{ML} techniques with synthetically created cyber attack data.

\subsection{Research Methodology} \label{subsec:investigation_procedure}
This research aimed to evaluate the influence of complex attack patterns on data integrity and to explore how varying levels of complexity impact the quality of data. For this purpose, we utilized a multi-layered network structure inspired by the Purdue model~\cite{b6}. This structure includes 21 sub-networks, each representing different facets of an industrial control power grid. We integrated diverse defensive strategies and offensive capabilities at various network levels. For instance, the spread of \ac{IDS} sensors within the corporate network was adjusted, which significantly affected the course of attack propagation. We found that deploying between 5 and 15 sensors optimally balanced the rate of generated alerts from suspicious activities against excessive interventions by defenders in our scenarios. We also varied parameters that control the attack’s speed and reach, facilitating the creation of complex data sets. The quality and variability of data were assessed by training a range of \ac{ML}-based anomaly detection models on attack logs formatted in the Unified2 \ac{IDS} event style~\cite{au2016multi}, as detailed in Table~\ref{tab:gen_alerts}. These logs served as inputs for our \ac{ML} models, designed to differentiate alerts resulting from regular operations or attacker-induced events.

The investigation encompassed synthetic data creation, game-theoretical modeling, optimizing attack paths, and constructing attack graphs with computational complexity of $O(N^3)$. 
\begin{table}[ht]
\centering
\caption{Generated Alerts from the tool in Unified2 format}
\label{tab:gen_alerts}
\begin{tabular}{llll}
\hline
label & size & label & size \\ \hline
sensor id & 4 bytes & source port/icmp type & 2 bytes \\
event id & 4 bytes & dest. port/icmp code & 2 bytes \\
event second & 4 bytes & protocol & 1 byte \\
event microsecond & 4 bytes & impact flag & 1 byte \\
signature id & 4 bytes & impact & 1 byte \\
generator id & 4 bytes & blocked & 1 byte \\
signature revision & 4 bytes & mpls label & 4 bytes \\
classification id & 4 bytes & vlan id & 2 bytes \\
priority id & 4 bytes & padding & 2 bytes \\
ip source & 16 bytes & application id & NA \\
ip destination & 16 bytes & sequence number & NA \\ \hline
\end{tabular}
\end{table}

\subsection{Examining Attack Complexity}
\begin{figure}[ht]
\centerline{\includegraphics[width=\columnwidth]{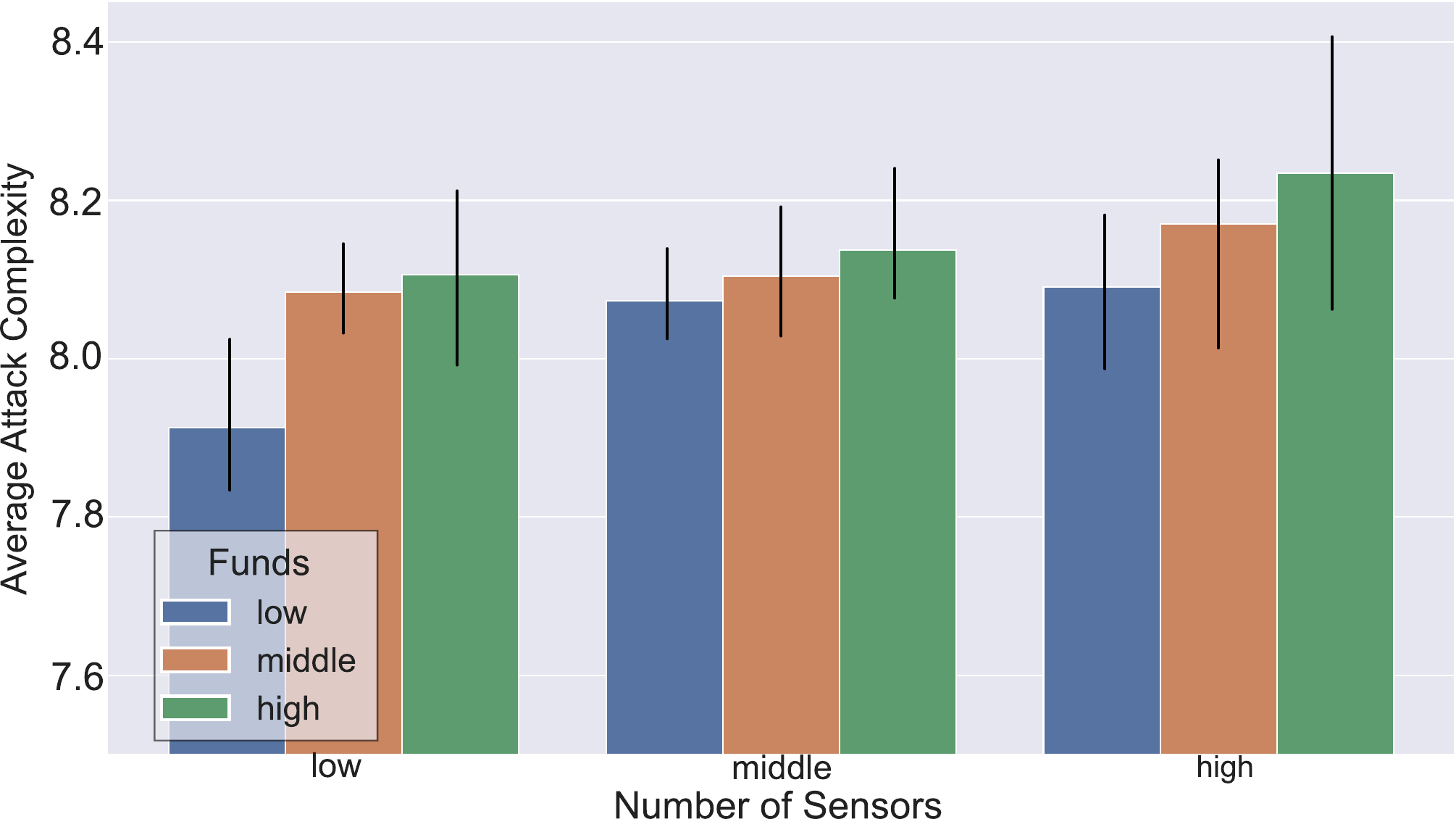}}
\caption{Complexity assessment of conducted attacks under various initial conditions. The y-axis denotes the mean complexity score throughout the simulations, complemented by a 95\% \ac{CI}. The x-axis shows the count of \ac{IDS} sensors deployed, while the bar colors indicate the level of defensive investment.}
\label{fig:attack_complexity}
\end{figure}

We assessed the variation in attack strategies by quantifying the complexity on a scale ranging from 0 (simplest) to 10 (highly varying), referencing the \ac{CVSS}~\cite{b25}, and analyzing the length of the attack propagation routes. To evaluate the \ac{ML} models effectively, we deployed them across 30 distinct attacks, using varying initial configurations (refer to Figure~\ref{fig:attack_complexity} for details). The attacker's expertise level was initially set at 0.5 and increased by 0.02 with each attack, eventually reaching a score of 1 by the 25th attack. We also experimented with three distinct financial scenarios concerning initial capital and incremental funds. Furthermore, the number of \ac{IDS} sensors was strategically altered to adjust the attack's difficulty level. Attack complexity was deduced based on the exploitation complexity of the vulnerabilities involved. As depicted in Figure~\ref{fig:attack_complexity}, the average complexity score for the vulnerabilities exploited in all 30 attacks is illustrated, complete with a 95\% \ac{CI}. Notably, a rise in the quantity of \ac{IDS} sensors and available funds correlated with an increase in attack complexity. Additionally, complexity escalated when the attacker had to navigate alternative routes following defensive countermeasures or exploit a greater number of vulnerabilities.

\subsection{Evaluating Model Performance}\label{subsec:performance}
For effective classification, we assessed various \ac{ML} algorithms as discussed in the \ac{IDS} literature review~\cite{b25} and supplemented by insights from another study~\cite{b26}. Among the supervised learning techniques, we opted for \ac{RF}, \ac{DT}, \ac{SVM}, \ac{CNB}, and \ac{XGB}. The choice of \ac{CNB} over its Gaussian counterpart was driven by its superior handling of imbalanced datasets, as noted in~\cite{b27}. For unsupervised learning, the K-Means algorithm was employed. Following the methodologies outlined in prior research~\cite{b25,b26}, we applied K-Means to time series data, fine-tuning the ‘k’ value through Hyper Parameter Grid Search, targeting optimal performance and reduced overfitting as observed in our experiments.

These models were tested iteratively against the generated data, utilizing the accumulated historical data for training and current data for evaluation. This approach ensured a careful balance between minimizing overfitting and maintaining predictability. Performance analyses were primarily conducted in scenarios with a moderate deployment of sensors and financial resources. While various scenarios were explored, the performance rankings of the models remained relatively consistent.

Table~\ref{tab:models_scores} showcases the efficacy of each \ac{ML} method in the final simulation iteration, evaluated using several classification metrics like accuracy, $F_{1}$-score, \ac{AUC}, and \ac{MCC}, particularly relevant for datasets with imbalances. A notable observation was the \ac{SVM}'s protracted processing time with large data volumes, aligning with the findings in~\cite{b28}, rendering it less suitable for our outlined evaluation method in Section~\ref{subsec:investigation_procedure}.

\begin{figure}[ht]
\centerline{\includegraphics[width=\columnwidth]{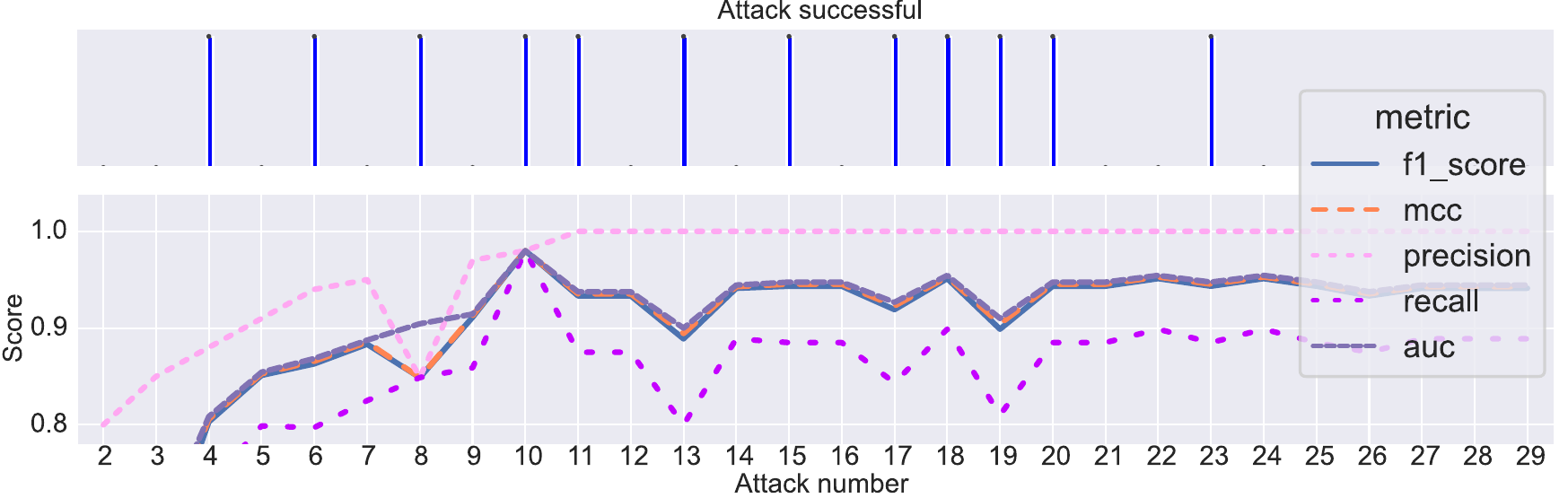}}
\caption{Evaluation of the \ac{RF} model.
The x-axis represents the simulation iteration, while the color labels of the lines represent different metrics.}
\label{fig:evaluation_rf}
\end{figure}

\begin{figure}[ht]
\centerline{\includegraphics[width=\columnwidth]{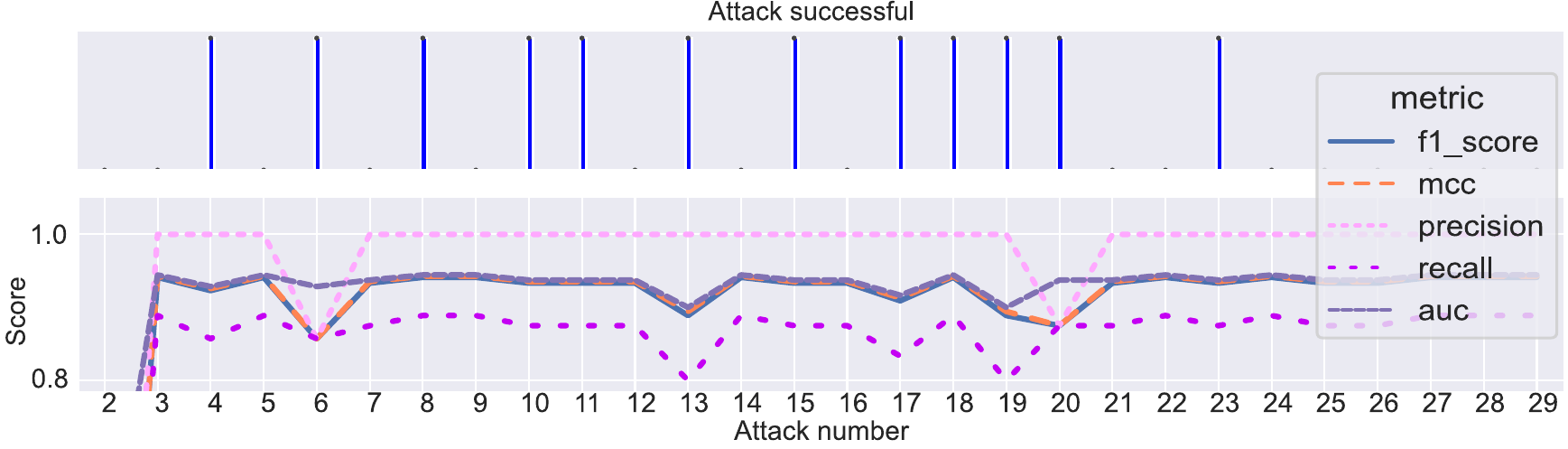}}
\caption{Evaluation of the \ac{XGB} model.
The x-axis represents the simulation iteration, while the color labels of the lines represent different metrics.}
\label{fig:evaluation_xgb}
\end{figure}

In our analyses, the K-Means algorithm lagged in performance, whereas the \ac{XGB} model excelled in both training efficiency and key performance metrics. Figures~\ref{fig:evaluation_rf} and \ref{fig:evaluation_xgb} illustrate the progression of the \ac{RF} and \ac{XGB} models across simulations 1-29, highlighting the correlation between the number of attacks and their respective success rates. Initially, both models faced challenges in accurately identifying specific attacks but gradually improved their understanding of attack dynamics. Interestingly, past the 23rd attack, no attacks were successful, and a skill plateau was achieved by the attacker by the 25th attack.

\begin{figure}[ht]
\centerline{\includegraphics[width=\columnwidth]{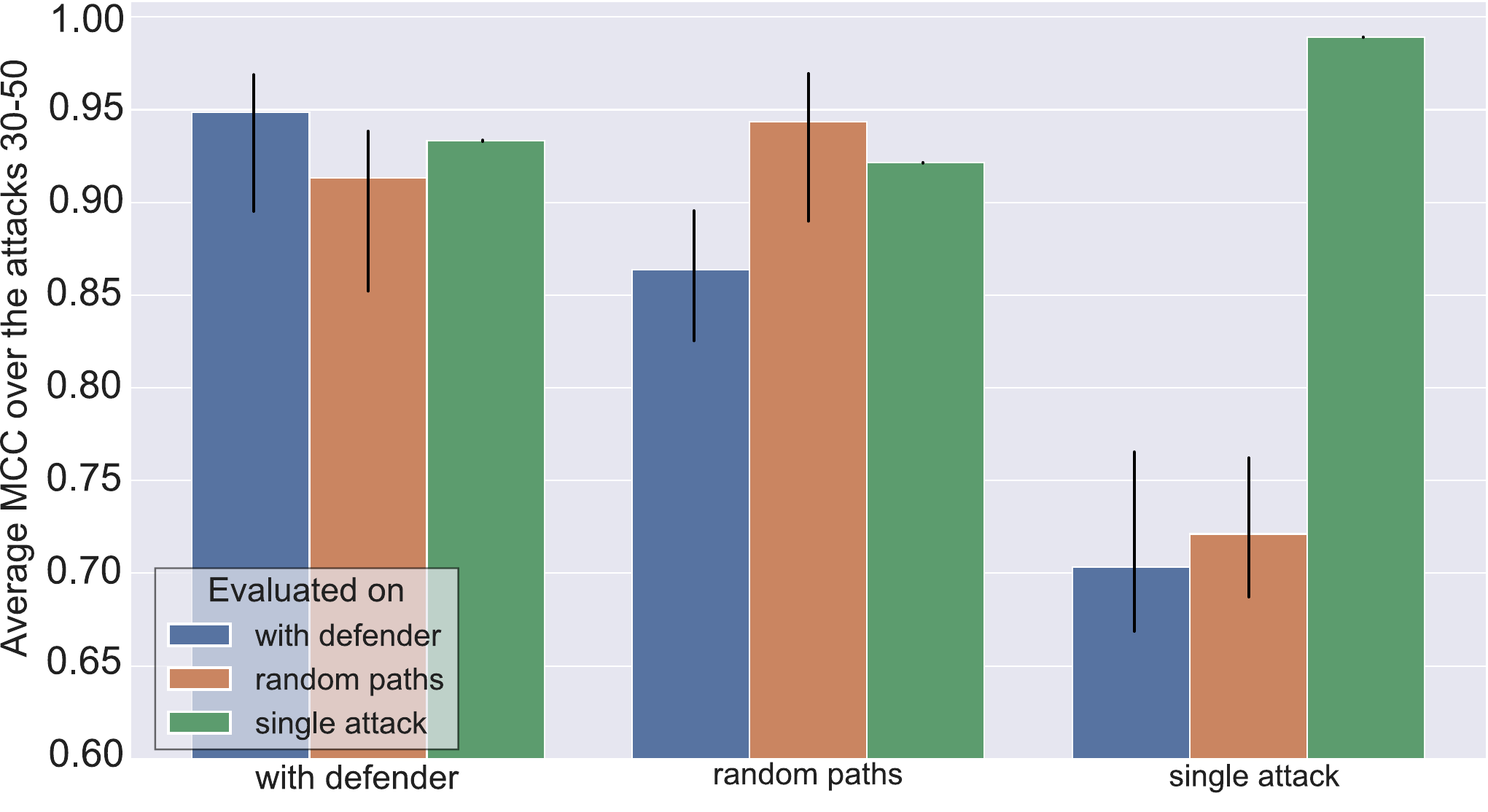}}
\caption{Evaluation of the different generation methods.
The y-axis represents the \ac{MCC} over simulation runs 30-50, while the error bar represents the 95\% \ac{CI}.
The x-axis represents the training data, and the color labels of the bars represent the test data.}
\label{fig:evaluation}
\end{figure}

Further, we assessed the effectiveness of the defender's role using three distinct attack data generation approaches (see Figure~\ref{fig:evaluation}). Extending the simulations to 50 runs ensured result consistency. The first approach involved interactions with the defender, as previously described. In the second approach, labeled ``single attack'', the attacker randomly navigated without defensive countermeasures. The third scenario saw the attacker choosing the most efficient path in the absence of any defense strategy. \ac{XGB} models, trained on data from the initial 29 attacks, were tested against the latter 21 attacks, each subjected to the various generation methods. Comparing the outcomes across ``random'', ``single attack'', and ``with defender'' scenarios demonstrated that data generated with an active defender role significantly enhanced detection quality. This improvement can be attributed to the diversified and realistic attack patterns emerging from the strategic modifications enforced by the defender, which were less apparent in the random or the single attack methodologies.

\begin{table}[ht]
\centering
\caption{Scores of the different \ac{ML} models using various metrics}
\begin{tabular}{|p{1.3cm}|p{0.75cm}|p{0.75cm}|p{0.75cm}|p{0.75cm}|p{0.75cm}|p{0.75cm}|}
\hline
Metric & \ac{RF} & \ac{DT} & \ac{SVM} & \ac{CNB} & K-Means & \ac{XGB} \\ \hline
Accuracy & 0.9375 & 0.8182 & 0.9382 & 0.6697 & 0.5003 & 1 \\ \hline
Recall & 0.8889 & 0.8889 & 0.89 & 0.6810 & 0.0003 & 0.8889 \\ \hline
Precision & 0.9999 & 0.7273 & 0.9889 & 0.7 & 0.5555 & 1 \\ \hline
$F_{1}$-Score & 0.9411 & 0.8000 & 0.9369 & 0.6904 & 0.0006 & 0.9412 \\ \hline
\ac{AUC} & 0.9444 & 0.8391 & 0.9251 & 0.6841 & 0.5279 & 0.9444 \\ \hline
\ac{MCC} & 0.9333 & 0.6471 & 0.8721 & 0.3367 & 0.0005 & 0.9428 \\ \hline
\end{tabular}
\label{tab:models_scores}
\end{table}

\subsection{Discussion}
Throughout our research, we noticed that an increase in both the number of deployed sensors and the available defense budget led to heightened complexity in attack patterns. This complexity emerged as attackers were compelled to navigate more challenging routes, often encountering sophisticated vulnerabilities. Such a shift in strategy might be attributed to either the original path being obstructed by the defender's sensors and reactive measures or the attacker's dwindling success rate owing to heightened preventative actions demanding more resources. Additionally, our analysis indicated a notable improvement in the \ac{ML} models' predictive capacities with an increase in the volume of attacks used for training, providing a richer data set for learning.

The continuous interaction between the attacker and defender created a varied training dataset, equipping the \ac{ML} models to better adapt and respond to evolving attack patterns. This was particularly apparent in scenarios where defender interventions were part of the training data, which resulted in superior detection capabilities. In our final set of experiments, it was evident that data reflecting the defender's participation in the attack scenarios established a more complex dynamic. In contrast, models trained on datasets devoid of this interactive aspect struggled to fully grasp the nuances of attack methodologies, underscoring the value of including defender-attacker interactions in training datasets for more effective modeling of diverse attack scenarios.

The increased complexity of attack patterns, driven by the defender’s proactive measures, illustrates the realistic and dynamic nature of the simulation, which not only replicates the evolving behavior of cyber attacks but also generates diverse datasets critical for training robust IDS models. The simulation’s ability to capture these interactions demonstrates how defense strategies directly influence the sophistication of attacks, thereby validating the importance of incorporating such dynamic scenarios into IDS training. This approach ensures that the \ac{IDS} models are better equipped to detect and respond to complex and evolving threats, highlighting the overall value of the simulation environment in advancing smart grid cyber security by offering a robust platform for both attack scenario generation and defense strategy evaluation. Thus, this interaction between attack and defense contributes to the broader goal of the paper: enhancing the resilience of smart grids through advanced simulation and data-driven defense mechanisms.

%% file: chapter4c.tex
\section{Decision Support System} \label{sec:dss}
Building upon the knowledge and experiences gleaned from the case studies, this section lays the foundation for introducing the \ac{DSS} for cyber security. The \ac{DSS} is characterized as a pivotal tool, developed with the objective of augmenting decision-making capabilities in the context of cyber threats. By harnessing data and findings derived from the simulated attack scenarios, the \ac{DSS} provides grid operators with a robust platform for strategic and informed decision-making within real-time cyber threat environments. This section also presents an exemplary application of the simulation environment, aimed at evaluating the effectiveness of the \ac{DSS} within the context of power grids.

\subsection{Design}\label{subsec:dss_proc}
The evolving landscape of cyber security, particularly in the realm of distribution grid management, necessitates a robust and adaptable \ac{DSS}. The system outlined here leverages \ac{ADT} to model threats and defenses~\cite{fila2020exploiting}. This methodical approach allows for a structured and comprehensive risk assessment, addressing the complex challenges faced by distribution grids.

\ac{ADT} are instrumental in visualizing and quantifying threats and defenses within distribution grids. These trees start with a broad attack category at the root, which is then dissected into more specific sub-attacks or countermeasures. The nodes within these trees are quantified with risk attributes---probability, impact, and cost---offering a detailed perspective on vulnerabilities and the efficacy of countermeasures.

After constructing the \acp{ADT} models, each node is quantitatively annotated with risk attributes using two principal equations: the product of probability and impact, $R_i = P_i \cdot I_i$, and a more comprehensive formula incorporating the cost of the attack, $R_i = \frac{P_i \cdot I_i}{C_i}$. This multifaceted risk assessment approach prevents overlooking the cost dimension, ensuring balanced and realistic threat evaluations.

The risk attributes assigned to each attack and defense node lead to a bottom-up propagation process. Here, the risk values from the leaf nodes are aggregated up to their parent nodes, culminating in a total risk assessment at the root node. This process reflects the overall risk posture of the system.

The Sobol method is utilized to analyze the variance-based global sensitivity of the system~\cite{chastaing2015generalized}. It quantifies how individual or grouped attributes contribute to overall risk variance, identifying which attributes most significantly influence the system's risk. This insight is critical for targeted resource allocation and prioritization.

The \ac{DSS} employs \ac{ADT} mincuts and a three-dimensional relational matrix for identifying cost-effective countermeasures~\cite{rios2020integrated}. This matrix helps in tackling single and multiple objective optimization problems, finding the minimal set of defenses that efficiently cover targeted attacks while minimizing costs.

Continuous risk assessment is a core feature of the system, crucial for adapting to the dynamic nature of cyber security threats. Regular updates to the probability and impact values of attack and defense nodes, informed by real-time security monitoring, ensure that the system's risk assessments are current and accurate.

The system also explores defense deployment optimization strategies, examining various combinations to achieve cost-effective risk mitigation. This involves considering factors such as return on investment and overall risk reduction.

Using risk quadrants aligned with the OWASP risk rating methodology~\cite{wiradarma2019risk}, the system categorizes attacks by severity and probability. This enables prioritizing defenses against the most significant risks, as evidenced by a marked reduction in threat density within the critical quadrant following countermeasure application.

\subsection{Experiment}\label{subsec:dss_res}
The \ac{DSS} within the simulated power grid environment was set up to evaluate the effectiveness of various countermeasures against a range of cyber security threats. The simulation was conducted to replicate a real-world scenario where an adversary's goal is to compromise the power grid system. The experiment was designed to test the robustness of defensive strategies and to assess their impact on mitigating the risks associated with different types of cyber attacks.

\begin{figure*}
\centerline{\includegraphics[width=\linewidth]{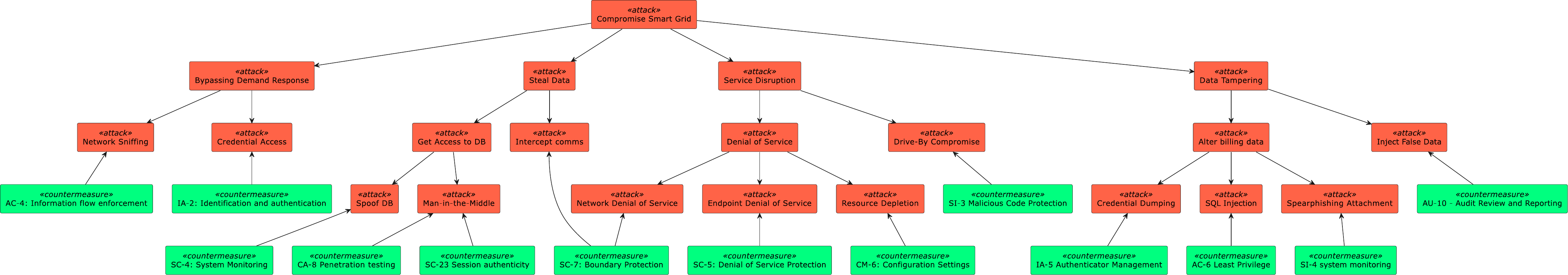}}
\caption{Illustration of the \ac{ADT} scenario for the \ac{DSS} experiment.}
\label{fig:dss_adt}
\end{figure*}

The simulation environment considers a structured \ac{ADT} (cf. Figure~\ref{fig:dss_adt}) with a hierarchical framework that mapped the progression from a broad threat objective to specific attack techniques. This structure included a root attack node, sub-objective attack nodes, intermediate attack nodes and strategies, leaf attack nodes, and corresponding countermeasure nodes.

\begin{figure}[ht]
\centerline{\includegraphics[width=\columnwidth]{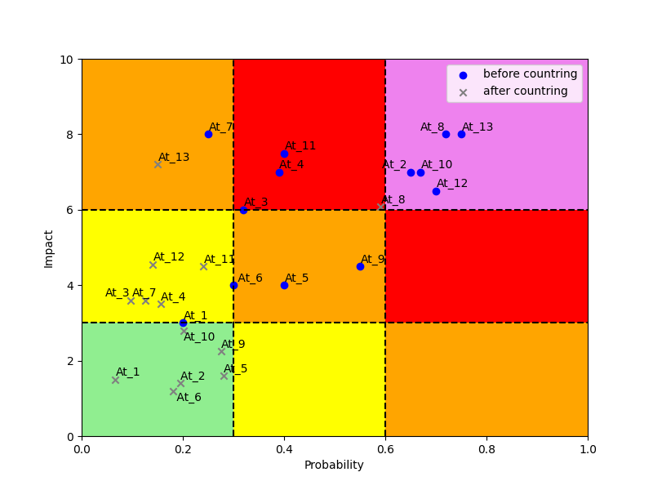}}
\caption{Illustration of quadrant plot depicting the \ac{DSS} result to the \ac{ADT} scenario.}
\label{fig:dss_quadrant}
\end{figure}

Upon execution of the initial risk assessment, the simulation provided risk values for each attack node, quantifying the probability and impact of potential cyber threats without any defensive measures (cf. Figure~\ref{fig:dss_quadrant}). These values served as a baseline to measure the effectiveness of the implemented countermeasures (cf. Table~\ref{tab:fused_risk_values}).

\begin{table}[H]
\centering
\caption{Risk vector values for attack events in ADT before and after bottom-up propagation}
\label{tab:fused_risk_values}
\begin{tabular}{|p{0.5cm}|p{2cm}|p{0.5cm}|p{0.3cm}|p{0.5cm}|p{0.5cm}|p{0.5cm}|p{0.5cm}|}
\hline
\textbf{Node Label} & \multicolumn{1}{c|}{\textbf{ADT Node Name}} & \textbf{Initial Probability} & \textbf{Initial Impact} & \textbf{Initial Risk} & \textbf{Upd. Probability} & \textbf{Upd. Impact} & \textbf{Upd. Risk} \\ \hline
At\_1 & Endpoint Denial of Service & 0.2 & 3 & 0.08 & 0.06 & 1.5 & 0.014 \\ \hline
At\_2 & Network Denial of Service & 0.65 & 7 & 0.75 & 0.19 & 1.4 & 0.04 \\ \hline
At\_3 & Resource Depletion & 0.32 & 6 & 0.48 & 0.09 & 3.6 & 0.08 \\ \hline
At\_4 & Drive-By Compromise & 0.39 & 7 & 0.68 & 0.15 & 3.5 & 0.13 \\ \hline
At\_5 & Network Sniffing & 0.4 & 4 & 0.22 & 0.28 & 1.6 & 0.06 \\ \hline
At\_6 & Credential Access & 0.3 & 4 & 0.17 & 0.18 & 1.2 & 0.03 \\ \hline
At\_7 & Intercept comms & 0.25 & 8 & 0.4 & 0.12 & 3.6 & 0.09 \\ \hline
At\_8 & Man in the middle & 0.72 & 8 & 1.15 & 0.59 & 6.08 & 0.71 \\ \hline
At\_9 & Spoof db & 0.55 & 4.5 & 0.61 & 0.27 & 2.25 & 0.15 \\ \hline
At\_10 & Spearphishing Attachment & 0.67 & 7 & 1.56 & 0.20 & 2.8 & 0.18 \\ \hline
At\_11 & Credential Dumping & 0.4 & 7.5 & 0.75 & 0.24 & 4.5 & 0.27 \\ \hline
At\_12 & SQL Injection & 0.7 & 6.5 & 0.60 & 0.14 & 4.55 & 0.08 \\ \hline
At\_13 & Inject false data & 0.75 & 8 & 1.5 & 0.15 & 7.2 & 0.27 \\ \hline
\end{tabular}
\end{table}

As countermeasures were introduced into the simulation, their effectiveness was determined by the updated risk values. The system measured the impact of these countermeasures on reducing the likelihood and potential damage of each attack scenario. Notable results included significant risk reduction for attacks such as ``Endpoint Denial of Service'' and ``Network Denial of Service'' where the introduction of countermeasures such as ``SC-5: Denial of Service Protection'' and ``SC-7: Boundary Protection'' lowered the risk from critical levels to less severe categories.

However, some attack vectors, such as \ac{MITM} remained resistant to the countermeasures, indicating a need for more sophisticated defensive strategies. This was consistent with the earlier observed note that even after applying defenses, At\_8 did not show a significant reduction in risk, implying that it did not have a substantial influence on the overall system risk.

A sensitivity analysis was conducted to understand the variability in the system's risk posture due to changes in the defense attributes. This analysis highlighted the importance of defense mechanisms such as ``Df\_4: Anomaly Detection Systems'' in minimizing the overall risk. However, it also identified a threshold beyond which further enhancements to certain countermeasures, such as ``Df\_4'' did not yield additional risk reduction, suggesting an optimal point for resource allocation.

The Sobol method used in the experiment underscored its suitability for complex, high-dimensional problems. It allowed for a rapid convergence, meaning that an accurate assessment of the system's risk could be achieved with fewer samples. This contributed to an efficient and comprehensive analysis of the power grid's cyber security posture.

\subsection{Discussion}\label{subsec:dss_disc}
The underlying simulation environment plays a crucial role in the development and testing of the \ac{DSS} for power grid cyber security. This environment, by its very design, is tailored to replicate the varying cyber physical behavior of power grid systems, along with the multifaceted aspects of cyber security impacts, encompassing various attack and defense mechanisms.

At the core of this simulation is the capacity to mimic real-world scenarios and cyber physical interactions within the power grid. This realism is essential for evaluating the system's resilience against a range of cyber threats and determining the efficacy of various defense strategies. By simulating the behavior of both the grid and potential cyber threats, the environment provides a comprehensive platform for analyzing how different attack vectors can affect the grid's functionality and reliability.

The \ac{DSS} emerges as a direct consequence of this simulated environment, drawing on the insights gained from these detailed analyses. Its development is predicated on the understanding that managing cyber security in power grids is not just about confronting isolated cyber incidents but involves a holistic approach to understanding and mitigating the cascading effects these incidents can have on the entire grid infrastructure.

In this context, the \ac{DSS} serves as an advanced tool that offers strategic guidance on incident response. It leverages the differences in modeling of \ac{ADT} to dissect and comprehend potential vulnerabilities and entry points within the power grid. This level of detail ensures that the strategies developed are both precise and tailored to the specificities of each threat.

Moreover, the simulation environment enables scenarios where a dynamic risk assessment process can be evaluated and tested. By continually adjusting to changing attack parameters and recalculating risk values, the \ac{DSS} exemplifies adaptability and sophistication. This constant evolution is crucial in a domain where threats are not static but evolve rapidly, requiring a response system that is equally agile and forward-looking.

The sensitivity analysis using the Sobol method, performed within this environment, enhances the \ac{DSS}'s capability to prioritize and allocate resources effectively. By understanding the systemic impact of various attack and defense attributes, the \ac{DSS} can guide decisions on where to focus efforts for maximum impact and cost efficiency.

However, it's essential to recognize that the simulation environment, while advanced, cannot perfectly replicate every real-world variable or unpredictable human factor in cyber attacks. As a result, while the \ac{DSS} provides invaluable insights and strategic guidance, its recommendations must be tempered with an understanding of its inherent limitations.

The simulation environment’s role in the development of the \ac{DSS} highlights the importance of a holistic and dynamic approach to power grid cyber security. By replicating different cyber-physical behavior, the simulation environment ensures that both attack and defense mechanisms can be comprehensively analyzed, providing a platform for realistic and detailed scenario generation. This realism is central to the effectiveness of the \ac{DSS}, which relies on accurate, real-time data to offer strategic incident response guidance. Furthermore, the dynamic risk assessment process that continuously evolves based on changing attack parameters underscores the adaptability required in modern cyber security solutions. The ability of the \ac{DSS} to recalibrate in response to varying attack vectors and defense strategies highlights the sophistication of the approach, aligning with the overall message of the paper: that robust, real-time simulation environments are essential for developing resilient and responsive cyber security systems for critical infrastructure. The sensitivity analysis and resource prioritization capabilities within the \ac{DSS} add further value, showing how simulation-driven insights can optimize defense strategies while acknowledging the system's limitations, such as its ability to replicate every unpredictable human element in real-world attacks. This reinforces the need for continued refinement and validation of both the simulation environment and the \ac{DSS} in evolving threat landscapes.

%% file: chapter5.tex
\section{Conclusion} \label{sec:conclusion}
\subsection{Summary} \label{subsec:conclusion_summary}
In this paper, we have developed and validated a novel, comprehensive simulation environment that integrates energy and communication systems into a single unified platform. This environment not only moves beyond the limitations of traditional co-simulation approaches but also offers significant contributions to the understanding and mitigation of cyber threats in power grids. Our model is capable of simulating complex cyber physical interactions, including sophisticated cyber attacks such as \ac{FDI}, and generating high-quality, synthetic datasets that are crucial for training \ac{ML}-based \ac{IDS}. These synthetic datasets provide researchers with a realistic yet flexible means to develop, train, and test \ac{IDS} in scenarios that are difficult or impossible to recreate in real-world environments due to the security, ethical, and logistical challenges associated with critical infrastructure.

A key contribution of our simulation environment lies in its ability to model the dynamic interaction between attackers and defenders, simulating real-time adjustments in defensive strategies based on the evolving nature of cyber threats. This dynamic aspect is crucial as static defense measures are often insufficient in the face of advanced, multi-stage cyber attacks. The dynamic simulation framework enables us to evaluate how real-time defensive measures, such as \ac{IDS} sensor placements and firewall configurations, impact the detection and mitigation of cyber threats. This approach provides new insights into the influence of defense strategies on overall network security and the effectiveness of \ac{ML}-based \ac{IDS}.

Moreover, the simulation environment provides a foundation for generating diverse datasets, a critical requirement for improving the adaptability and accuracy of ML models used in \ac{IDS}. By simulating attack scenarios with varying complexities, we were able to produce data that reflects different attack vectors, from simple one-step attacks to complex, multi-stage attacks involving multiple entry points into the power grid's communication and operational technology layers. The ability to simulate these variations ensures that the datasets used for training \ac{IDS} are sufficiently diverse, preparing the system to detect a wide range of threats. This is particularly important because real-world cyber attacks often evolve and adapt in response to the defender's strategies. The ability of the environment to simulate both attacker adaptation and defender responses ensures that \ac{IDS} models trained in this environment are robust and capable of detecting even previously unseen attack strategies.

The results of our work also emphasize the importance of parameterization in cyber attack simulations. The interaction between firewall configurations, \ac{IDS} sensor placements, and the metadata associated with the attacker significantly influences the vulnerability landscape of the simulated network. For instance, variations in the placement of \ac{IDS} sensors across different sub-networks and the corresponding firewall rules resulted in significant differences in the detection capabilities of the \ac{IDS} models. These findings underscore the importance of fine-tuning the defensive infrastructure within power grids to maximize the effectiveness of detection and response systems.

The validation of our simulation environment was carried out through a series of laboratory tests, in which we replicated cyber attack scenarios both in the simulation environment and in a real-world cyber physical laboratory setup. The results of these experiments demonstrated the high precision of our simulation platform in replicating both the physical behavior of the power grid and the communication processes under both normal and attack-induced conditions. By comparing the results of the simulation with those of the real-world system, we were able to assess the accuracy of the simulated environment in terms of replicating real-world scenarios. This validation process is a critical step in establishing the credibility of the simulation as a tool for \ac{IDS} and DSS development, as it ensures that the synthetic data generated for training \ac{IDS} models is both realistic and comprehensive.

A significant aspect of the validation process involved the comparison of grid state estimations under normal and attack conditions. The replication of \ac{FDI} attacks within both environments highlighted the vulnerability of grid state estimations to manipulations. These manipulations can lead to severe consequences, such as incorrect decision-making by grid operators, resulting in widespread outages or damage to the grid infrastructure. The simulation environment's ability to replicate these scenarios underscores the importance of robust detection mechanisms, such as \ac{ML}-based \ac{IDS}, in safeguarding grid operations from cyber threats.

In addition to providing a platform for \ac{IDS} development, our simulation environment also supports the development and testing of a \ac{DSS} for power grid cyber security. The \ac{DSS} synthesizes data from the \ac{IDS} and other monitoring systems to offer real-time strategic guidance during cyber incidents. This capability is particularly important in the context of smart grids, where quick decision-making is critical to preventing or mitigating the effects of cyber attacks. The ability of the \ac{DSS} to provide actionable insights in real time allows grid operators to respond to incidents more effectively, thereby reducing the potential impact of an attack. The dynamic nature of the simulation environment, which includes real-time adjustments to defense strategies and attack progression, ensures that the \ac{DSS} is trained on data that reflects realistic cyber threat scenarios, further enhancing its effectiveness.

This paper presents a sophisticated and fully integrated simulation environment that advances the state of the art in smart grid cyber security. By providing a platform for generating realistic training data, validating attack scenarios, and supporting dynamic defense strategies, this environment serves as a powerful tool for both \ac{IDS} and \ac{DSS} development. The contributions of this work lie not only in the development of the simulation environment itself but also in the insights gained from the dynamic interaction between attackers and defenders, the importance of parameterization, and the role of validation through laboratory testing. While our simulation environment is robust, it relies on theoretical models that may not fully capture the complexity of real-world human factors and cyber attacks. Additionally, the scope of the platform, although comprehensive in simulating cyber-physical interactions in power grids, could be expanded further.

\subsection{Outlook} \label{subsec:conclusion_outlook}
Future work should aim to scale the simulation to include larger and more complex grid scenarios, including \acp{DER} and microgrids. Expanding the scope of the simulation to include these additional elements will provide a more comprehensive view of the potential vulnerabilities within modern power grids, as well as the strategies that can be employed to defend against cyber threats. Additionally, the incorporation of human factors into the simulation environment, such as the role of human operators in responding to cyber incidents, would provide a more complete perspective on power grid cyber security.

A further area for improvement is the incorporation of more advanced algorithms within both the \ac{IDS} and \ac{DSS}. As cyber threats continue to evolve, so too must the tools used to detect and respond to them. Incorporating state-of-the-art ML algorithms into the \ac{IDS}, as well as more advanced decision-making algorithms into the \ac{DSS}, will help ensure that these systems remain effective in the face of increasingly sophisticated cyber attacks. In particular, the use of reinforcement learning techniques, which allow systems to learn and adapt over time based on feedback from previous incidents, could significantly improve the adaptability of both the \ac{IDS} and \ac{DSS}.

Moreover, there is potential to extend the use of synthetic data generation beyond \ac{IDS} training to other areas of cyber security research. For example, the data generated by our simulation environment could be used to develop new algorithms for automated incident response, or to evaluate the effectiveness of different network configurations in mitigating the impact of cyber attacks. By leveraging the flexibility of synthetic data, researchers can explore a wide range of scenarios and strategies that would be difficult or impossible to investigate in a real-world setting.

Future work will focus on expanding the scope of the simulation, incorporating more advanced algorithms, and addressing the role of human factors in cyber attack resilience. These developments will further enhance the robustness and effectiveness of power grid cyber security solutions, helping to protect critical infrastructure from the growing threat of cyber attacks.